\documentclass[10pt,final,doublecolumn]{IEEEtran}
\usepackage{amsthm}
\usepackage{amsmath}
\usepackage{latexsym}
\usepackage{graphicx}
\usepackage{bbding}
\usepackage{indentfirst}
\usepackage{algorithm,algorithmic}
\usepackage{setspace}
\usepackage{float}
\usepackage{epstopdf}
\usepackage{amssymb}
\usepackage{amsfonts}
\usepackage{enumerate}
\usepackage{multicol}
\usepackage{color}
\usepackage{bm}
\usepackage{amssymb}
\usepackage{stfloats}
\usepackage{epstopdf}
\usepackage{threeparttable}
\usepackage{epstopdf}
\usepackage{threeparttable}
\usepackage{makecell}
\usepackage{cite}
\usepackage{graphicx}
\usepackage{subcaption}
\usepackage[labelsep=period]{caption}  

\IEEEoverridecommandlockouts
\allowdisplaybreaks[4]

\def\BibTeX{{\rm B\kern-.05em{\sc i\kern-.025em b}\kern-.08em
		T\kern-.1667eb\lower.7ex\hbox{E}\kern-.125emX}}
\usepackage{balance}
\begin{document}
	\title{Flexible Coupler Array with Reconfigurable Pattern: Mechanical Beamforming and Digital Agent}
	\author{{Xiaodan Shao, \IEEEmembership{Member, IEEE}, Yixiao Zhang, Nan Cheng, \IEEEmembership{Senior Member, IEEE}, Weihua Zhuang, \IEEEmembership{Fellow, IEEE}, Xuemin (Sherman) Shen, \IEEEmembership{Fellow, IEEE}}
		\thanks{X. Shao, Y. Zhang, W. Zhuang, and X. Shen are with the Department of Electrical and Computer Engineering, University of Waterloo, Waterloo, ON N2L 3G1, Canada (E-mail: \{x6shao, y3549zha, wzhuang, sshen\}@uwaterloo.ca)}
		\thanks{Nan Cheng is with the School
			of Telecommunications Engineering, Xidian University, Xi’an
			710071, China (e-mail: 
			dr.nan.cheng@ieee.org).}
}
		
	\maketitle

\begin{abstract}
Flexible coupler is a promising solution for enhancing wireless network capacity by moving passive couplers around a fixed-position active antenna to reshape the induced currents on passive elements.
Motivated by this, this paper proposes a novel flexible coupler array that incorporates additional degrees of freedom (DoF) in radiation pattern reconfiguration and enhanced communication coverage with low hardware cost.
Specifically, a new form of mechanical beamforming can be obtained
by moving only the passive coupling elements while keeping
the active antenna stationary. In addition, the flexible coupler
antenna can slide along a rail toward users, thereby enhancing communication coverage. To fully exploit the potential of the flexible coupler
array, we formulate a two-timescale sum-rate maximization
problem with statistical channel state information (CSI). The antenna position is optimized
based on scattering cluster-core statistics in the slow timescale,
while mechanical beamforming is optimized based on multipath
channel statistics in the fast timescale, subject to movement and
energy constraints. To address the coupling between timescales
and the high cost of extensive channel sampling, we develop a
digital agent framework that leverages an electromagnetic (EM)
map to generate statistical channel information for different
user and antenna positions. Then, a deep neural network is
trained to learn a slow-fast performance (SFP) surrogate, which
is fine-tuned with a small number of real measurements and
then applied for position optimization at the slow timescale
using projected gradient ascent. Mechanical beamforming at
the fast timescale is obtained by selecting per-antenna radiation
patterns from a predefined dictionary via a convex relaxation.
Simulation results demonstrate that the proposed flexible coupler
array significantly improves system throughput, and the digital agent-assisted algorithm achieves satisfactory performance with greatly reduced online computational complexity.
\end{abstract}

		
	\section{Introduction}
	Multiple-input multiple-output (MIMO) communication technology will remain pivotal for
	future sixth-generation (6G) and beyond, where the demand for massive connectivity, ultra-low latency, and
	high energy efficiency is driving the exploration of advanced
	wireless communication technologies. The MIMO technology achieves
	these goals through spatial multiplexing, high beamforming
	gains, and multi-user communications \cite{exl,Larsson2014Massive,8937497}. An appealing feature of MIMO is that its performance gains scale with the number of antennas. As MIMO dimensions increase, however, it inevitably amplifies critical system-level challenges, including steadily increasing signal processing overhead, hardware cost, energy consumption, and computational complexity.
Moreover, since the antennas in conventional MIMO systems are deployed at fixed positions, the system lacks the capability to adaptively reconfigure or optimize the wireless channel within the transmit-receive region in response to the user distribution and the environmental dynamics.
	Consequently, communication systems in 6G and beyond will require more efficient transceiver architectures to support a tenfold or greater increase in MIMO dimensions.
		To enhance wireless system performance, several new MIMO technologies have emerged. For example, by intelligently adjusting the phase shifts of the reflecting elements, the reconfigurable intelligent surface (RIS)/intelligent reflecting surface (IRS) can dynamically reconfigure the transceivers' effective channel gains \cite{proc,9724202,10443321}. Holographic MIMO surfaces realize large continuous. Yet, it is desirable to further enable flexible allocation of antenna resources in accordance with the spatially nonuniform distribution of users, to enhance their performance gains.

To fully exploit spatial channel variations, the six-dimensional movable antenna (6DMA) has recently been proposed to increase MIMO system capacity without requiring additional antennas \cite{shao20246d,6dma_dis,10945745}. This technique leverages the adaptability of antenna positions and rotations (i.e., orientations) in three-dimensional (3D) space at transceivers to adaptively allocate antenna resources based on the spatial distribution of channels. By providing improved array, spatial multiplexing, and geometric gains while effectively suppressing interference, the 6DMA system aims at enhancing wireless network performance.
In \cite{10989638,passive6DMA}, the position and rotation of a passive IRS are exploited to enable flexible adjustment of the beamforming direction. In \cite{6DMA_JSTSP}, low-training-complexity channel estimation algorithms are proposed by exploiting a new directional-sparsity characteristic of 6DMA channels.
In \cite{jiang2025statistical}, a low-complexity sequential optimization approach was proposed that first determines 6DMA rotations and then finds feasible positions to realize the optimized rotations subject to practical antenna position constraints. Then, a 6DMA-enabled wideband terahertz (THz) communication system with a sub-connected hybrid beamforming architecture was studied in \cite{yan2025six}. The 6D movable holographic-surface-based integrated data and energy transfer system was examined in \cite{wang20256d} and \cite{shen20256d}, where the holographic surface or metasurface is both translatable and rotatable within a predefined space. The authors in \cite{liu2024uav} and \cite{wen} investigated a new unmanned aerial vehicle (UAV)-enabled passive 6DMA architecture by mounting an IRS on a UAV and addressed the associated joint deployment and beamforming optimization problem. In \cite{gaohierarchical}, an enhanced 6DMA-assisted over-the-air computation system was proposed to minimize the computation mean square error in Internet of Things (IoT) networks. 
Moreover, the authors in \cite{shao2025tutorial} discussed two special cases of 6DMA, namely a rotatable antenna with fixed position and a translatable antenna with fixed rotation.
Other studies also explored cell-free 6DMA \cite{free6DMA,pi20246d}, 6DMA-enhanced wireless sensing \cite{6dmasensing}, polarized 6DMA \cite{IPA}, hybrid near-far field 6DMA\cite{near}, and channel knowledge map-enabled 6DMA \cite{11314850}. In addition, fluid antennas, movable antennas, and pinching antennas can adjust antenna positions, thereby enhancing communication performance \cite{10436574,MA,zeng2025robust}.

   \begin{figure*}[!t]
	\centering
	\includegraphics[width=0.73\linewidth]{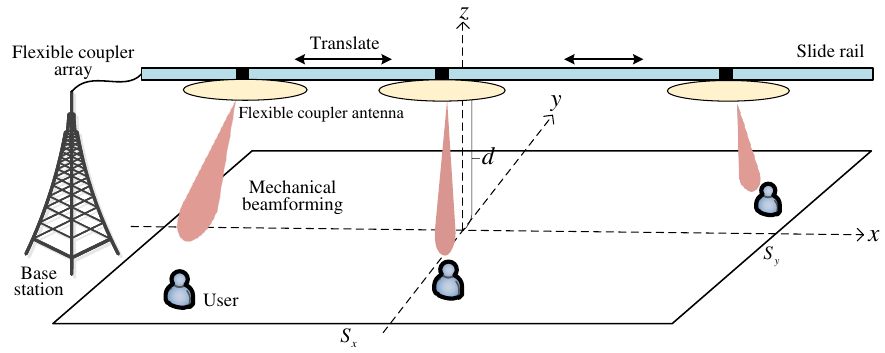}
	\caption{Illustration of flexible coupler array for mitigating large-scale path loss and enabling mechanical beamforming.}
	\label{EMA}
	\vspace{-0.59cm}
\end{figure*} 

Despite its promising applications, the existing 6DMA designs move the active antenna element together with its radio-frequency (RF) feed over a sufficiently large region with a size on the order of several to tens of wavelengths to exploit spatial channel variations, which introduces substantial mechanical complexity and energy consumption. Fortunately, the authors in \cite{Shao2026FC} proposed a low-complexity flexible coupler antenna, where mutual coupling can serve as an enabler of advanced spatial processing to improve system performance. By dispensing with the requirement that each antenna be connected to a dedicated RF chain, each flexible coupler antenna is connected to a single RF chain and multiple passive couplers \cite{Shao2026FC}. By rotating or shifting only passive couplers within a small area via electromagnetic (EM) coupling, the induced coupling currents can be reshaped to enhance communication performance without moving the active antenna or the associated RF hardware \cite{Shao2026FC}. However, existing flexible coupler antennas typically permit coupler translations on the order of several wavelengths. Such constraints hinder the mitigation of large-scale path loss and consequently restrict communication performance.


Moreover, existing 6DMA systems assume directional antennas with fixed radiation patterns. Such fixed patterns limit the ability of the antennas to adapt to spatial channel variations and to shape the radiated energy according to the environment. In contrast, by mechanically translating/rotating passive couplers and tuning their geometries, the EM radiation pattern can be reconfigured, which introduces an additional radiation DoF for each antenna element and improves the exploitation of channel diversity \cite{Shao2026FC}. In addition, optimization and channel estimation for 6DMA mainly rely on real-world data measurements. As the number of antennas and antenna reconfigurable parameters increases, exhaustive testing in physical environments becomes infeasible given the ultra-low latency requirements of future wireless systems. 
To address this challenge, we introduce digital agent, which is conceptually related to that of the digital twin \cite{11025157,11214475,10638530, 10643661,10623528,9906937,10886938}. It replicates physical assets in a digital space for testing and analysis without interfering with the real system.
Unlike digital twins in manufacturing \cite{Grieves2014}, which primarily focus on monitoring and offline simulation, an antenna digital agent emphasizes modeling network behavior, learning the characteristics of the radio environment, and enabling real-time configuration as well as customized network management. As a practical implementation of a digital agent, the EM map, also referred to as the channel knowledge map, has been proposed \cite{10287775,9373011,10764739}. It is a site-specific database tagged with the locations of transmitters and receivers, containing channel-related information useful to enhance environment-awareness and facilitate or even obviate sophisticated real-time channel state information (CSI) acquisition. Unlike conventional channel knowledge maps that aim to predict instantaneous channel realizations, the EM map considered in this paper stores slowly varying statistical channel information derived from environmental geometry and user positions, which reduces the overhead of channel reconstruction. In this way, the digital agent can provide reliable statistical CSI and performance surrogates under stringent latency constraints, thereby supporting efficient antenna position optimization and mechanical beamforming in flexible coupler systems.
 
	The aforementioned challenges motivate the study of flexible coupler array with reconfigurable pattern in this work. The main contributions of this paper are summarized as follows. 
	\begin{itemize}
		\item 
		We propose a flexible coupler array in which passive couplers move around each active antenna to realize mechanical beamforming for radiation-pattern reconfiguration. The flexible coupler antenna can also slide along a rail to move close to the receiver, thereby establishing a strong communication link and mitigating large-scale path loss. Mechanical beamforming enables a reconfigurable radiation pattern that extends angular coverage, which reduces the number of required slide rails and the hardware cost for achieving full spatial coverage. Moreover, moving passive couplers instead of the active antenna avoids moving RF hardware, thus providing fine beam-steering capability while reducing mechanical complexity.
	
	\item  We formulate an optimization problem to maximize the user sum rate by jointly designing the mechanical beamformer and the antenna position vector based on channel statistics. To solve the problem, we develop a digital agent-assisted two-timescale optimization framework for flexible coupler array. At the fast timescale, the mechanical beamformer is optimized to maximize the ergodic sum rate under statistical  CSI by convex relaxation. At the slow timescale, we extract statistical CSI from an EM map to train a slow-fast performance (SFP) surrogate with a deep neural network for predicting the optimized sum rate as a function of the antenna position vector, which already accounts for the fast-timescale mechanical-beamforming optimization. The surrogate is further fine-tuned online and used to update antenna position by a single projected gradient step. The proposed design avoids repeated fast-timescale mechanical beamforming during position optimization, reduces online complexity, and removes the need for explicit prior knowledge of the instantaneous channel.
	
\item
	Extensive simulations are conducted to validate the capability of the proposed flexible coupler array to enhance system performance, as well as the effectiveness of the proposed digital-agent-aided two-timescale algorithm in reducing online computational complexity and adapting to changing environments.
	\end{itemize}
	
In practice, the proposed flexible coupler array with digital agent can be deployed in different applications. For example, a slide rail can be mounted along the wall of a subway station \cite{8468019}, where the digital agent uses the EM map to track user flows along the platform and adapt antenna positions and radiation patterns to overcome strong blockage by pillars, thereby maintaining coverage. In indoor hotspots such as offices, shopping malls, and residential buildings, the proposed array can move antennas closer to rooms with heavy traffic and steer their patterns around walls and furniture \cite{moon2022deep}, which can improve service quality. Similar deployments are feasible in factories \cite{lutzmayr2022wireless}, where the digital agent updates the EM map when machinery/shelves move and then reoptimizes the mechanical beamforming so that the network maintains reliable throughput.
	
The rest of this paper is organized as follows. Section II presents the flexible coupler array together with the corresponding channel and signal models. Section III introduces the slow  and fast timescale protocol for the flexible coupler system and formulates the sum rate maximization problem by jointly designing mechanical beamforming and antenna positions under practical constraints. Section IV describes the digital agent-aided algorithm for antenna position and mechanical beamforming design. Section V discusses numerical results for performance evaluation and comparison. Finally, Section VI concludes this study.

\emph{Notations}: Boldface upper-case and lower-case letters denote
matrices and vectors, respectively, $(\cdot)^H$ and $(\cdot)^T$  respectively denote conjugate transpose and transpose, and $\mathbb{E}[\cdot]$ denotes the expected value of
	a random variable. For a scalar $a$, $|a|$ denotes its magnitude. 
	For a vector $\mathbf{a}$, $\|\mathbf{a}\|$ and $\|\mathbf{a}\|_1$ denote the $\ell_2$-norm and $\ell_1$-norm, respectively.  Operator $\mathrm{diag}({\bf x})$ denotes a diagonal matrix with the diagonal entries specified by vector ${\bf x}$, $[\mathbf{a}]_j$ denotes the $j$-th element of vector $\mathbf{a}$, $[\mathbf{A}]_{i,j}$ denotes the element of matrix $\mathbf{A}$ at the $i$-th row and $j$-th column, $\otimes$ denotes the Kronecker product, $\mathbf{I}_M$  denotes an $M \times M$ identity matrix, $\mathrm{blkdiag}\{\mathbf{A}_1, \mathbf{A}_2, \ldots, \mathbf{A}_M\}$ denotes a block diagonal matrix, $\mathrm{vec}(\mathbf{A})$ denotes the vectorization operator that stacks all the columns	of a matrix $\mathbf{A}$ into a single column vector,
	 $\mathcal{O}(\cdot)$ denotes the big-O notation, $\mathbb{R}$ and $\mathbb{C}$ denote the real and complex fields, respectively, $\mathcal{N}(\mu,\sigma^2)$ denotes a real Gaussian distribution with mean $\mu$ and variance $\sigma^2$, and 
	$\mathcal{CN}(\boldsymbol{\mu},\boldsymbol{\Sigma})$ denotes a circularly symmetric complex Gaussian distribution with mean $\boldsymbol{\mu}$ and covariance matrix $\boldsymbol{\Sigma}$.
		
	\section{System Model}
	
\begin{figure}[!t]
		\centering
		\includegraphics[width=0.59\linewidth]{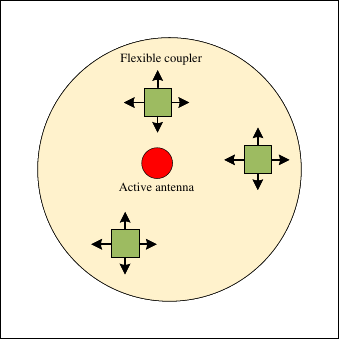}
		\caption{Top view of the flexible coupler antenna with reconfigurable antenna pattern.}
		\label{C6DMA}
			\vspace{-0.59cm}
	\end{figure}

We consider an uplink system where a base station (BS) serves a set of single-antenna users $\mathcal{K}=\{1,\ldots,K\}$, with $K$ denoting the total number of users.
 As illustrated in Fig.~\ref{EMA}, the BS is equipped with a flexible coupler array, where $N$ flexible coupler antenna are mounted on one slide rail, and each flexible coupler antenna can move along the slide rail on a scale much larger than a wavelength and can be positioned close to users to form a strong communication link. The slide rail is installed at a fixed height of $d$ and aligned parallel to the $x$-axis. As shown in Fig.~\ref{C6DMA}, each flexible coupler antenna consists of one fixed-position active antenna and multiple passive couplers that move within a designated area around the active antenna  \cite{Shao2026FC}. The positions of the passive couplers can be mechanically adjusted with the aid of drive components such as micro-electromechanical systems (MEMS), which provide low power consumption and fast response  \cite{Shao2026FC}. The active antenna is connected to a single RF chain, and the receiver is assumed to have a fixed single antenna. By moving the passive coupler around the fixed active radiator, the antenna’s radiation pattern is reconfigured through the induced currents on the passive element without moving the active antenna’s RF chains \cite{6691942}.
	The users are distributed within a known two-dimensional area, e.g., a rectangular region with a size of $S_x \times S_y~\mathrm{m}^2$. The location of the active antenna in the $n$-th flexible coupler antenna on the slide rail in the Cartesian coordinate system is given by $[p_n, 0, d]$, where $0 \le p_n \le X$ denotes the position of the active antenna on a slide rail of length $X$.

\subsection{Mechanical Beamforming}
We refer to the capability of steering and shaping the EM radiation pattern by mechanically relocating the passive coupler elements around the fixed active antenna as mechanical beamforming  \cite{Shao2026FC}. In mechanical beamforming, the beam direction and beam shape are controlled through the geometry and positions of the passive couplers, while the RF chains of the active antenna remain fixed.

\begin{figure}[!t]
		\centering
		\includegraphics[width=0.83\linewidth]{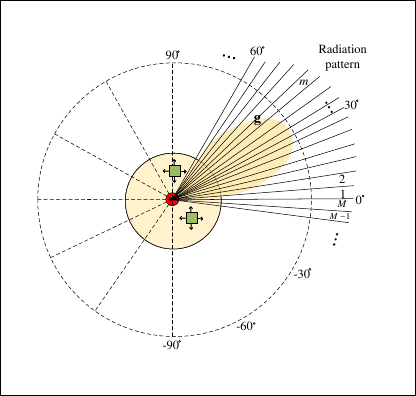}
		\caption{Reconfigurable antenna pattern for flexible coupler antenna.}
		\label{radition_pattern}
		\vspace{-0.59cm}
	\end{figure}

To quantify the performance gains enabled by mechanical beamforming with flexible couplers, we establish a mathematical model for the radiation pattern and the associated channel characteristics. The flexible coupler introduces an additional DoF in the angular domain, which expands the information coverage of a conventional antenna from a one-dimensional spatial representation to an $M$-dimensional angular representation, where $M$ is the number of uniformly sampled directions. The reconfigurable radiation pattern in the angular domain is therefore characterized over $M$ discrete angular directions (see Fig. \ref{radition_pattern}).
To model reception characteristics, we introduce the virtual angular-index channel for any flexible coupler antenna, denoted as 
	\begin{align}
		\label{eq:hem}
		\mathbf z &\in \{0,1\}^M,
	\end{align}
	which serves as an index vector indicating the spatial angles at which the user is located. The radiation pattern for any flexible coupler antenna in the EM  angular domain is represented by $\mathbf g\in \mathbb R^M$, with each element corresponding to the radiation gain in one of the $M$ discrete directions. Consequently, the effective gain in the EM  angular domain can be written as
	\begin{align}
		B
		= \mathbf z^T \mathbf g,
	\end{align}
	which quantifies the radiation intensity directed toward the user's actual spatial angle. By optimizing the radiation pattern $\mathbf g$ to align with the angular‐domain channel characteristics $\mathbf z$, their correlation and thus the effective radiation efficiency, $B$, is maximized. This formulation serves as the foundation for the following theoretical analysis and optimization.  

	Building upon the radiation pattern model of a single flexible coupler antenna presented above, we extend this model to characterize the programmable radiation pattern of the entire flexible coupler array at the BS, which consists of \(N\) flexible coupler antennas. The resulting mechanical beamforming (also called radiation beamforming) in the EM  angular domain is expressed in the following matrix form
	\begin{align}
		\mathbf G
		&\triangleq \mathrm{blkdiag}\bigl\{\mathbf g_{1},\,\mathbf g_{2},\,\dots,\,\mathbf g_{N}\bigr\}\in\mathbb R^{M N\times N},
		\label{eq:FEM}
	\end{align}
	where the \(m\)-th element of \(\mathbf g_{n}\) represents the radiation gain of the \(n\)-th antenna in the \(m\)-th angular direction.  To ensure fairness in pattern design, the total energy of each antenna’s radiation pattern is normalized as \(\|\mathbf g_{n}\|^2=1,\;\forall n\).
	
	To further characterize the spatial mapping of the flexible couplers, we introduce the virtual angular‐index channel 
	\begin{subequations}
		\begin{align}
			\bar{\mathbf Z}_{k}
			&\triangleq 
			\mathrm{blkdiag}\{\mathbf{Z}_{k,1}, \ldots, \mathbf{Z}_{k,N}\} 
			\in \{0,1\}^{MN \times L_kN},
			\label{eq:bigHEM}\\
			\mathbf Z_{k,n}
			&\triangleq \bigl[\mathbf z_{k,n,1},\,\mathbf z_{k,n,2},\dots,\mathbf z_{k,n,L_k}\bigr]
			\in\{0,1\}^{M\times L_k},
			\label{eq:smallHEM}
		\end{align}
	\end{subequations}
	where \(\mathbf Z_{k,n}\) specifies the spatial angles associated with the $n$-th antenna and the \(k\)-th user.  Each column \(\mathbf z_{k,n,l}\in\{0,1\}^{M}\) contains exactly one nonzero entry, indicating the angular direction of the \(l\)-th $(l\in \{1,2,\cdots,L_k\})$ propagation path on the $n$-th antenna for user \(k\), with $L_k$ being the number of propagation paths.  Consequently, radiation gain \(\mathbf B_{k}\in\mathbb R^{L_k N\times N}\) of all antennas toward the \(k\)-th user can be formulated as
	\begin{align}
		\mathbf B_{k}
		&=\bar{\mathbf Z}_{k}^T\,\mathbf G=\mathrm{blkdiag}\bigl\{\mathbf Z_{k,1}^T\mathbf g_{1},\,\dots,\,\mathbf Z_{k,N}^T\mathbf g_{N}\bigr\}.
		\label{eq:GEMk}
	\end{align}
	
	In summary, matrices \(\mathbf G\) and \(\bar{\mathbf Z}_{k}\) mathematically capture the additional DoF introduced by flexible coupler array, enabling a more flexible and adaptive radiation beamforming strategy in the EM  angular domain. Note that antenna position vector $\mathbf{p}=[p_{1},p_{2},\cdots,p_{N}]^T$ and mechanical beamforming vector $\mathbf{G}$ are the design parameters that we tune to optimize the flexible coupler system throughput.
	
	\textit{Remark 1:} 
	Unlike prior 6DMA designs that steer beams by moving the \emph{active} antenna \cite{shao20246d}, the proposed flexible coupler array keeps the active antenna fixed and instead moves a lightweight passive coupler in the reactive near field. Induced currents on the passive element reshape the radiation pattern, so there is no need to move the active antenna’s RF hardware or add switching networks, yielding fine angular resolution with much lower mechanical and RF complexity \cite{Shao2026FC}.
	
	\subsection{Channel Model}	
Due to the sparse scattering characteristics of wireless channels at high frequencies, e.g., millimeter-wave (mmWave) bands, the channel typically consists of a few dominant clusters, each containing multiple propagation paths \cite{8122055}. Let $C_k$ denote the total number of scattering clusters for user $k$, and $L_{c,k}$ denote the number of multipaths in cluster $c$ for user $k$. Then, the near-field channel between the $n$-th flexible coupler antenna and the $k$-th user can be expressed as
\begin{align}
	h_{k,n}(p_{n})
	&= \sum_{c=1}^{C_k} \sum_{l_c=1}^{L_{c,k}}
	\frac{\alpha_{k,c,l_c}}{r_{c,l_c,n}(p_{n}) \, r_{k,c,l_c}}
	e^{-j \frac{2\pi}{\lambda}\big(r_{c,l_c,n}(p_{n})+r_{k,c,l_c}\big)},
	\label{lala}
\end{align}
where $\lambda$ denotes the carrier wavelength, $\alpha_{k,c,l_c}\in\mathbb{C}$ is the complex scattering coefficient associated with the $(c,l_c)$-th path for user $k$, $r_{k,c,l_c}$ is the distance from user $k$ to the $l_c$-th scatterer in cluster $c$, and $r_{c,l_c,n}(p_n)$ is the distance from that scatterer to the $n$-th flexible coupler antenna located at position $p_n$ on the sliding rail. 

The distance, $r_{c,l_c,n}$, can be described as
\begin{align}
	r_{c,l_c,n} = \bar{r}_{c,n} + \Delta r_{l_c,n}, \label{rr0}
\end{align}
where $\bar{r}_{c,n}$ denotes the nominal distance from the $c$-th cluster center to the $n$-th candidate position on the slide rail, and $\Delta r_{l_c,n}$ denotes the deviation of the $l_c$-th ray from the cluster nominal center, following a Gaussian distribution
\begin{align}\label{88}
	\Delta r_{l_c,n} \sim \mathcal{N}(0, \varsigma_{n,c}^{2}),
\end{align}
with $\varsigma_{n,c}^{2}$ indicating the distance spread.
Similarly, $r_{k,c,l_c}$ can be described as
\begin{align}
	r_{k,c,l_c} = \bar{r}_{k,c} + \Delta r_{k, l_c}, \label{rr1}
\end{align}
where $\bar{r}_{k,c}$ denotes the nominal user-cluster distance for the $c$-th scattering cluster and user $k$, and $\Delta r_{k, l_c}$ denotes the corresponding deviation, following a Gaussian distribution
\begin{align}\label{99}
	\Delta r_{k, l_c} \sim \mathcal{N}(0, \iota_{k,c}^{2}),
\end{align}
with $\iota_{k,c}^{2}$ indicating the distance spread.

For clarity, we define $
	L_k \triangleq \sum_{c=1}^{C_k} L_{c,k}
$
as the total number of paths. With this notation, the near-field spatial channel coefficient between the $n$-th flexible coupler antenna and the $k$-th user in \eqref{lala} can be reformulated as
\begin{align}
	h_{k,n}(p_{n})
	&= \sum_{l=1}^{L_k}
	\frac{\alpha_{k,l}}{r_{l,n}(p_{n}) \, r_{k,l}}
	e^{-j \frac{2\pi}{\lambda}\big(r_{l,n}(p_{n})+r_{k,l}\big)}.
	\label{eq:hkmn}
\end{align}

Furthermore, the integration of the reconfigurable pattern extends the channel representation from a one-dimensional to an $M$-dimensional formulation by incorporating additional radiation channel information, i.e., $\bar{\mathbf Z}_{k}$ in \eqref{eq:bigHEM}. The dimension-extended channel for the $n$-th flexible coupler antenna, denoted by 
$\mathbf{h}_{k,n}(p_n) \in \mathbb{C}^M$, is given by
\begin{align}
	\mathbf h_{k,n}(p_n)
	&=\sum_{l=1}^{L_k}
	\frac{\alpha_{k,l}}{r_{l,n}(p_{n}) \, r_{k,l}}
	e^{-j \frac{2\pi}{\lambda}\big(r_{l,n}(p_{n})+r_{k,l}\big)}  \mathbf z_{k,n,l}.
	\label{teq:hk_n_EM}
\end{align}
By stacking the channel vectors $\mathbf h_{k,n}(p_n)\in\mathbb{C}^{M}$ of all $N$ antennas, we obtain the complete channel vector for user $k$ as follows
\begin{subequations}
	\label{8}
	\begin{align}
		\mathbf h_k (\mathbf p)
		&=[\mathbf{h}_{k,1}^T(p_1), \cdots,
			\mathbf{h}_{k,N}^T(p_N)
]^T \\
	&	= \bar{\mathbf Z}_{k}\,\bar{\mathbf h}_{k}(\mathbf p)\in \mathbb{C}^{MN},
		\label{thf}
	\end{align}
\end{subequations}
where $\bar{\mathbf Z}_{k}$ is defined in \eqref{eq:bigHEM} and
\begin{align}\label{fo}
	\bar{\mathbf h}_{k}(\mathbf p)=\mathrm{vec}\big({\mathbf {A}}_{k}^{T}(\mathbf p)\big)\in\mathbb{C}^{NL_k},
\end{align}
with $
{\mathbf {A}}_k(\mathbf p)=
[
\alpha_{k,1}\,\mathbf a_{k,1}(\mathbf p),~\dots,~
\alpha_{k,L_k}\,\mathbf a_{k,L_k}(\mathbf p)]\in\mathbb{C}^{N\times L_k}$ and
$\mathbf a_{k,l}(\mathbf p)
	=[
	\frac{e^{-j \frac{2\pi}{\lambda}(r_{l,1}(p_{1})+r_{k,l})}}{r_{l,1}(p_{1})\,r_{k,l}},
	~\dots,~
	\frac{e^{-j \frac{2\pi}{\lambda}(r_{l,N}(p_{N})+r_{k,l})}}{r_{l,N}(p_{N})\,r_{k,l}}
]^{T}$.

This additional positional flexibility on a scale much larger than a wavelength enables the proposed flexible coupler array to establish a stronger and more stable communication link to the user than conventional fixed-antenna systems, which lack the ability to dynamically reconfigure or reconstruct the wireless channel.

\subsection{Signal Model}
We consider the uplink transmission from $K$ single-antenna users to the BS. The received signals at the BS are given by
\begin{align}
	\mathbf{y} = \mathbf{G}^H\mathbf{H}(\mathbf{p})\mathbf{x} + \mathbf{n}, \label{eq:uplink_model}
\end{align}
with
\begin{align} \label{98}
\mathbf{H}(\mathbf{p}) = [\mathbf{h}_1(\mathbf{p}), \mathbf{h}_2(\mathbf{p}), \cdots, \mathbf{h}_K(\mathbf{p})] \in \mathbb{C}^{MN\times K},
\end{align}
denoting the multiple-access channel from all $K$ users to the flexible coupler array at the BS.
In \eqref{eq:uplink_model}, $\mathbf{x} = \sqrt{\rho}[x_1, x_2, \cdots, x_K]^T \in \mathbb{C}^{K\times 1}$ with $x_k$ denoting the transmit signal of user $k$ with the average power normalized to one, and $\rho$ representing the transmit power of each user (assumed to be identical for all users). Vector 
$\mathbf{n} \sim \mathcal{CN}(\mathbf{0}, \sigma^2\mathbf{I}_{N})$ denotes the complex additive white Gaussian noise (AWGN) vector at the BS with zero mean and average power $\sigma^2$.

According to Shannon's
formula, the achievable sum-rate of all users over unit radio
spectrum bandwidth is given by 
\begin{align}
R(\mathbf{p},\mathbf G) &= \log_2 \det \left( \mathbf{I}_{N} + \frac{1}{\sigma^2} \sum_{k=1}^K \rho \mathbf{G}^H\mathbf{h}_k(\mathbf{p}) \mathbf{h}_k^H(\mathbf{p})\mathbf{G} \right) \notag\\
	&= \log_2 \det \left( \mathbf{I}_{N} + \frac{\rho}{\sigma^2} \mathbf{G}^H\mathbf{H}(\mathbf{p}) \mathbf{H}^H(\mathbf{p})\mathbf{G} \right), \label{r}
\end{align}
in bits per second per Hertz (bps/Hz). It is worth noting that, in contrast to the conventional multiuser channel with fixed-position antennas, the capacity of the flexible coupler-enabled wireless channel in \eqref{r} is determined by antenna positions $\mathbf{p}$ and mechanical beamforming (i.e., radiation beamforming) $\mathbf G$.

\section{Problem Formulation }
\subsection{Two Timescale Protocol with Statistical CSI}
	\begin{figure*}[!t]
	\centering
	\includegraphics[width=1.0\linewidth]{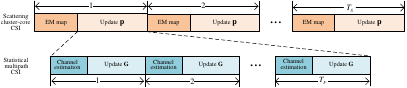}
	\caption{Proposed slow  and fast timescale protocol for flexible coupler array.}
	\label{cprotocol}
		\vspace{-0.59cm}
\end{figure*} 

The proposed flexible coupler array encounters practical challenges, including high channel estimation overhead and considerable movement complexity. Specifically, joint optimization of antenna positions and mechanical beamforming based on instantaneous CSI requires high-speed and frequent mechanically controlled translations, which incur substantial implementation cost and complexity. Moreover, the adjustment speed of such mechanical movements is inherently slow, so the movement duration may exceed the coherence time of the instantaneous channel, which makes instantaneous-CSI-based designs impractical. 

To address this limitation, we adopt statistical CSI, under which each flexible coupler antenna moves much more slowly and less frequently to adapt to long-term channel statistics. Even under statistical CSI, the timescales of translation and mechanical beamforming must be treated hierarchically. Antenna positions involve mechanical movements over distances much larger than the wavelength, typically from a few to several tens of meters, which limits their update speed and makes them suitable for adapting to slow-level variations, where the scattering cluster core remains nearly constant over extended periods. In contrast, the mechanical beamforming of the flexible coupler array is enabled by MEMS \cite{6691942} that move within a small range on the order of the wavelength, which allow rapid position adjustments to track fast-level variations, where the intra-cluster multipath statistics remain approximately stationary. Motivated by these considerations, we propose a hierarchical transmission framework that operates on scattering cluster-core statistics in the slow timescale and multipath channel statistics in the fast timescale to balance implementation feasibility and system performance, as illustrated in Fig.~\ref{cprotocol}.
\begin{itemize}
\item \textbf{Slow timescale:} The considered time frame, \(T\), is partitioned into $T_S$ super-frames, denoted by $\mathcal{T}_S = \{1,2,\ldots,T_S\}$. In each super-frame, the BS extracts cluster-core statistics (using the EM map discussed in Section IV) and optimizes the antenna position vector to adapt coverage to the dominant scattering-core geometry.
	
	\item \textbf{Fast timescale:} Each super-frame is further divided into $T_F$ frames, defined as $\mathcal{T}_F = \{1,2,\ldots,T_F\}$. In each frame, the BS acquires a statistical multipath channel  within the scattering clusters to update the mechanical beamformer for enhancing array gain.
\end{itemize}

This hierarchical framework is to reduce channel estimation overhead while ensuring efficient position and mechanical beamforming adaptation across different timescales. 

\subsection{Problem Formulation}
Based on the flexible coupler array,
we propose to maximize the achievable sum-rate of users by jointly designing  mechanical beamformer $\mathbf G$ in the fast timescale and antenna position vector $\mathbf{p}$ in the slow timescale, while satisfying the constraints of radiation pattern energy and position constraints. Thus, the design is formulated as
\begin{subequations}\label{MG3}
	\begin{align}
		\text{(P1)}\;
		&\max_{\mathbf p, \mathbf G} ~\mathbb{E}\Bigl[
		R(\mathbf p,\mathbf G)
		\Bigr] \label{gl}\\
		\text{s.t. }\;
		&\|\mathbf g_n\|^2 = 1,\quad \forall n, \label{M00}\\
		& 0\le p_n\le X, \forall n, \label{z81}\\
			& |p_n-p_{n'}|\ge d_{\min},\;\forall n\neq n', \label{z82}
	\end{align}
\end{subequations}
where 
constraint \eqref{M00} normalizes the radiation pattern energy of each antenna to ensure fairness across antenna elements.
Constraint \eqref{z81} confines antenna positions within the physical limits of the slide rail.
Constraint \eqref{z82} enforces a minimum separation between antennas to avoid collisions and excessive mutual coupling. 
In \eqref{gl}, $\mathbb{E}[\cdot]$ denotes the expectation over multipath channel realizations at the fast timescale, conditioned on the scattering cluster-core statistics at the slow timescale.
To address problem (P1), we need to overcome the following two challenges.
\begin{enumerate}
	\item \emph{Limited real data in high-dimensional antenna configurations.}
	Collecting user-level channels for every position-radiation combination of a flexible coupler array is prohibitively time-consuming and costly. A lightweight digital agent is therefore needed to provide reliable statistical CSI and performance surrogates within real-time constraints.
	
	\item \emph{Cross-timescale coupling between position and mechanical beamforming.}
	When a conventional two-timescale optimization scheme \cite{9198125} is used to solve problem (P1), each slow-timescale update triggers a full re-execution of the fast-timescale inner loop, which creates nested outer-inner iterations. The overall complexity thus scales rapidly with the number of antennas, leading to heavy computational load and substantial online latency.
\end{enumerate}

In the following subsections, we present a novel digital-agent-based antenna algorithm that enables intelligent optimization of antenna position and mechanical beamforming, thereby addressing the two challenges.

\section{Antenna Position and Mechanical Beamforming Design via Digital Agent}
In this section, we first describe the design of fast timescale mechanical beamforming, and then elaborate on the construction of the antenna digital agent for slow timescale antenna position vector optimization.

\subsection{Fast Timescale Mechanical Beamforming}
At the fast timescale, mechanical beamformer $\mathbf G$ mainly aims to establish a basic antenna pattern alignment with the statistical multipath channel.
In each fast timescale frame, the BS first estimates the statistical multipath channel given antenna-position vector $\mathbf p$, and then obtains the fast timescale mechanical beamforming matrix by solving
\begin{subequations}
	\label{MG33}
	\begin{align}
		\text{(P2)}\quad&\mathop{\max}\limits_{\mathbf G}\;
			\tilde{R}(\mathbf G)\\
		\text{s.t.}\quad&\|\mathbf g_n\|^2=1,\ \forall n, \label{8M00}
	\end{align}
\end{subequations}
where $\tilde{R}(\mathbf G)=\mathbb{E}[{R}(\mathbf G)]\triangleq \frac{1}{Z_{\mathrm{s}}}\sum_{t=1}^{Z_{\mathrm{s}}} R\big(\mathbf G;{\mathbf H}^{(t)}\big)$
denotes the ergodic sum-rate evaluated over $Z_{\mathrm{s}}$ fast-timescale channel samples
${\mathbf H}^{(t)}\in\mathbb C^{MN\times K}$ given in \eqref{98}, which are generated based on statistical CSI. The statistical CSI is obtained during uplink channel estimation, where all users simultaneously transmit pilot signals to the BS, and the BS applies the space alternating generalized expectation maximization (SAGE) algorithm to extract the channel statistics \cite{753729}.

Considering the hardware implementability of mechanically controlled flexible couplers, we adopt a practical approach by selecting the optimal radiation pattern from a predefined set, rather than designing arbitrary patterns with unrestricted directions and shapes. This approach ensures a balance between implementation efficiency and real-world hardware constraints. To be specific, the mechanical beamforming matrix is reformulated as
\begin{align}\label{eq:23}
	\mathbf{G}
=
	{\mathbf Q}
	\mathbf S,
\end{align}
where
$
{\mathbf Q}
=
\mathbf I_{N}\otimes\overline{\mathbf Q}
\in\mathbb R^{MN\times UN}
$ with $\overline{\mathbf Q}\in\mathbb R^{M	\times U}$
denotes a predefined dictionary that contains the radiation patterns generated under 
$U$ candidate patterns over 
$M$ spatial sampling angles. The matrix
\begin{align}
\mathbf S
\triangleq
\mathrm{blkdiag}\{\mathbf s_{1},\mathbf s_{2},\dots,\mathbf s_{N}\}
\in\{0,1\}^{UN\times N},
\end{align}
denotes the radiation pattern-selection matrix, where each
\(\mathbf s_{n}\in\{0,1\}^U\) selects one feasible position state of coupler (i.e., radiation pattern) from the dictionary, i.e., \(\|\mathbf s_{n}\|_1=1\). By substituting \eqref{eq:23} into \eqref{r}, we have
\begin{subequations}
\begin{align}
	\tilde{R}(\mathbf S) &=\mathbb{E}\left[\log_2 \det \left( \mathbf{I}_{N} + \frac{\rho}{\sigma^2} 
	\mathbf S^H{\mathbf Q}^H\mathbf{H}(\mathbf{p}) \mathbf{H}^H(\mathbf{p}){\mathbf Q}
	\mathbf S \right)\right]\\
	&=	\mathbb{E}\left[\log_2 \det \left( \mathbf{I}_{K} + \frac{\rho}{\sigma^2} \mathbf{H}^H(\mathbf{p}){\mathbf Q}
	\mathbf S 
	\mathbf S^H{\mathbf Q}^H\mathbf{H}(\mathbf{p}) \right)\right],
	\label{rrrs}
\end{align}
\end{subequations}
where \eqref{rrrs} is obtained based on Sylvester's determinant theorem, i.e., $\det(\mathbf{I} + \mathbf{A}\mathbf{B}) = \det(\mathbf{I} + \mathbf{B}\mathbf{A})$ \cite{akritas1996various}.

Based on the formulation in \eqref{rrrs}, the original problem in (P2) is equivalently transformed into a more practical optimization problem focused on optimizing the radiation pattern-selection matrix \(\mathbf S\), given by
\begin{subequations}\label{eq:24}
	\begin{align}
		\max_{\mathbf S}\;&\tilde{R}(\mathbf S) \label{eq:24a}\\
		\text{s.t. }\;&[\mathbf s_{n}]_u\in\{0,1\},\ \forall n,u, \label{eq:24b}\\
		&\|\mathbf s_{n}\|_{1}=1,\ \forall n. \label{eq:24c}
	\end{align}
\end{subequations}
The non-convex optimization problem in \eqref{eq:24} is particularly challenging because  the Boolean constraint \eqref{eq:24b} is both non-smooth and non-convex. The optimal selection algorithm is an exhaustive search over
all possible radiation pattern combinations. However, since the total number of feasible pattern selections is $U^{N}$, exhaustive search is impractical due to the extremely large number of possible antenna pattern combinations. Hence, a more tractable optimization strategy is required.
Specifically, for each fast-timescale channel sample, we let
\begin{align}
	\mathbf W_t &= (\mathbf{H}^{(t)}(\mathbf{p}))^H \mathbf Q\in\mathbb C^{K\times UN}. \label{fko}
\end{align}
Moreover, since $\mathbf{S}$ is a block diagonal matrix and each $\mathbf{s}_n$ is a binary indicator vector with exactly one unity entry, the product $\mathbf{S}\mathbf{S}^H$ becomes a diagonal selection matrix. Its diagonal is given by \(\mathbf v \triangleq [\mathbf s_1^{T}, \ldots, \mathbf s_N^{T}]^{T} \in \{0,1\}^{UN}\). Therefore, we obtain
\begin{align}
	\mathrm{diag}(\mathbf v) = \mathbf S 
	\mathbf S^H\in\mathbb C^{UN\times UN}. \label{fo1}
\end{align}
By substituting \eqref{fko} and \eqref{fo1} into \eqref{rrrs}, the ergodic sum-rate reduces to
\begin{align}
	\tilde R(\mathbf v) 
	&= \frac{1}{Z_{\mathrm{s}}}\sum_{t=1}^{Z_{\mathrm{s}}}\log_2\det\!\Big(\mathbf I_{K} + \tfrac{\rho}{\sigma^2}\,\mathbf W_t\mathrm{diag}(\mathbf v) \mathbf W_t^H\Big).
\end{align}
We then optimize the binary selector, $\mathbf v\in\{0,1\}^{UN}$, to maximize $\tilde R(\mathbf v)$ subject to per-antenna selection constraints.
Since the variables in $\mathbf v$ are binary integer, the optimization problem becomes NP-hard. 
To overcome this challenge, we relax the binary constraint on each element of $\mathbf v$ to a weaker continuous constraint such that $[\mathbf v]_i \in [0,1]$. 
The original problem in \eqref{eq:24} can be relaxed to the following problem.
\begin{subequations}\label{fff}
	\begin{align}
		\max_{\mathbf v}\;&\frac{1}{Z_{\mathrm{s}}}\sum_{t=1}^{Z_{\mathrm{s}}}\log_2\det\!\Big(\mathbf I_{K} + \tfrac{\rho}{\sigma^2}\,\mathbf W_t\mathrm{diag}(\mathbf v) \mathbf W_t^H\Big) \label{fv1}\\
		\text{s.t. }&\sum_{u=1}^{U} [\mathbf v]_{\phi(n,u)} = 1,~ n=1,\ldots,N,\\
		&0\le [\mathbf v]_i\le 1, ~i=1,\ldots,UN.
	\end{align}
\end{subequations}
where $\phi(n,u)=u+(n-1)U$. Problem \eqref{fff} is a convex optimization problem that can be solved in polynomial time using CVX \cite{cvx}.

After solving problem \eqref{fff} and obtaining the optimal relaxed solution $\mathbf v^*$,
we recover a feasible per-antenna binary selection vector  by groupwise rounding. Specifically, for each flexible coupler antenna $n$, we have
\begin{align}
	 \chi_n^* &= \arg\max_{u\in\{1,\ldots,U\}}\big[\mathbf v^*\big]_{\phi(n,u)},~n=1,\ldots,N, \label{pp0}\\
	\mathbf s_n &= \mathbf e_{ \chi_n^*}\in\{0,1\}^{U},~n=1,\ldots,N, \label{pp1}
\end{align}
where $\mathbf e_{u}$ denotes the $u$-th canonical basis vector in $\mathbb R^{U}$.
With the optimal radiation pattern selection vector $\{\mathbf s_n\}$ and known radiation pattern dictionary $\mathbf Q$, the mechanical beamformer is constructed as 
$\mathbf{G} ={\mathbf Q} \mathbf S$ in \eqref{eq:23}.

The computational complexity of solving problem (P2) in each fast timescale mainly depends on the solution of problem \eqref{fff}, whose complexity is of order $\mathcal{O}\!( \sqrt{2UN}\,\log (\tfrac{1}{\xi}) ( (UN)^3 + Z_{\mathrm{s}}K^2UN + Z_{\mathrm{s}}K^3 ) )$, where $\xi$ denotes the accuracy tolerance of the CVX solver.

\subsection{Digital Agent for Slow Timescale Position Optimization}
\begin{figure*}[!t]
	\centering
	\includegraphics[width=1.01\linewidth]{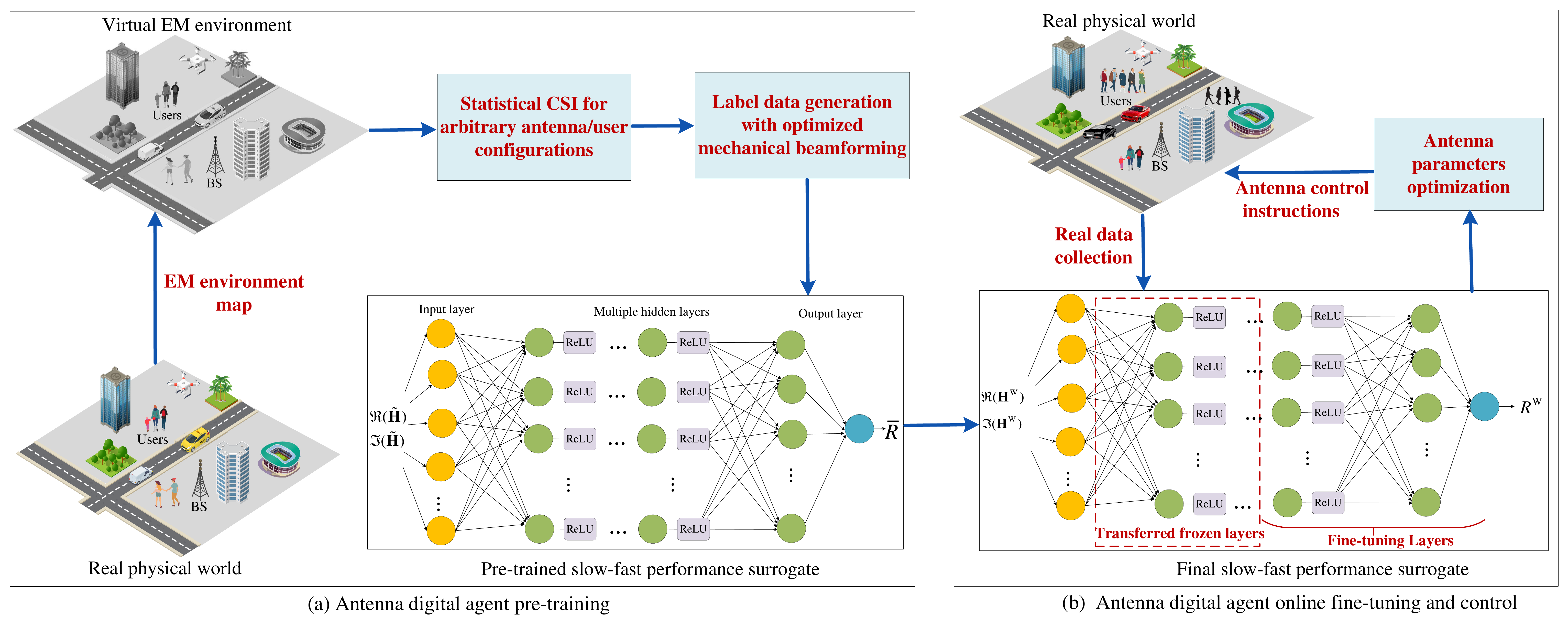}
	\caption{The proposed antenna digital agent architecture.}
	\label{learning}
		\vspace{-0.59cm}
\end{figure*} 

Over the slow timescale, antenna position vector $\mathbf p$ is optimized to place the array in the vicinity of the scattering-cluster cores and users, thereby strengthening the large-scale channel links. This design relies only on cluster-core information, namely the estimated nominal distances from users/antennas to the cluster cores. Accordingly, the cluster-core channel, $\mathbf h_{\mathrm{C},k}(\mathbf p)$, is defined as
\begin{align}
	\mathbf h_{\mathrm{C},k}(\mathbf p)
	= \bar{\mathbf Z}_{\mathrm{C},k}\,\bar{\mathbf h}_{\mathrm{C},k}(\mathbf p)\in\mathbb C^{MN},
	\label{or}
\end{align}
where $\bar{\mathbf h}_{\mathrm{C},k}(\mathbf p)\in\mathbb C^{C_kN}$ collects the spatial-domain channel components contributed by the $C_k$ scattering-cluster cores, and
$\bar{\mathbf Z}_{\mathrm{C},k}\in\{0,1\}^{MN\times C_kN}$ is a binary mapping matrix that encodes the angle information of the scattering cluster cores.
Consequently, we substitute the instantaneous multipath channel, $\mathbf h_k(\mathbf p)$, in the channel expression \eqref{thf} and the sum-rate formulation \eqref{r} with the scattering-cluster-core channel $\mathbf h_{\mathrm{C},k}(\mathbf p)$, thereby defining the slow-timescale sum rate $\tilde{R}(\mathbf p)$ as our optimization objective. Given mechanical beamformer $\mathbf G$, the slow timescale antenna position design problem is then formulated as
\begin{subequations}\label{p33}
	\begin{align}
		\text{(P3)}\;
		&\max_{\mathbf p}\;
		 \tilde{R}(\mathbf p)\\[2pt]
		\text{s.t. }\;
		& 0\le p_n\le X, \forall n, \label{bt85}\\
		& |p_n-p_{n'}|\ge d_{\min},\;\forall n\neq n'. \label{bt86}
	\end{align}
\end{subequations}

When optimizing the antenna position vector in (P3), deriving closed-form expressions of the
achievable sum rates of all users, $\tilde{R}(\mathbf p)$, in terms of the antenna position vector $\mathbf{p}$ only is difficult, because we are unable to obtain the optimal mechanical beamforming matrices as explicit functions of the antenna position vectors. As the number of antennas and the number of candidate positions increase, nested iterations over the resulting large parameter space under the strict latency requirements of future wireless links become infeasible for conventional control loops.
To make the problem tractable, we propose an SFP surrogate training approach with a digital agent, where the surrogate of the objective function is constructed based on appropriately generated channel realizations/samples and the current antenna position vector. 

An overview of antenna digital agent-based wireless systems is given in Fig. \ref{learning}. The framework leverages the EM map to generate statistical channel information across different antenna positions and user locations. Based on this synthetic dataset, a deep neural network is trained to learn an SFP surrogate that captures the coupling between antenna position and mechanical beamforming. To enhance generalization, the surrogate is further fine-tuned using a limited number of real channel measurements. This digital-agent-assisted algorithm enables efficient position optimization at the slow timescale and mechanical beamforming design at the fast timescale, while avoiding the nested iterations required in conventional two-timescale algorithms.
The proposed antenna digital agent comprises the following steps.

\subsubsection{EM Map}
According to \eqref{8}, besides antenna position vector $\mathbf{p}$, the wireless channel is determined by the environmental model, the user position distribution, and the propagation operator, which captures how EM waves interact with the medium and surrounding objects. In an ideal implementation, the agent would maintain up-to-date information on all dynamic scatterers. However, real-time mapping of every movable antenna is often impractical due to cost, power, and computational constraints. Since large-scale propagation parameters are determined by the environmental geometry and user positions, path loss and angular characteristics exhibit slow temporal variation. This makes it feasible to pre-store a set of statistical information within the EM map (also referred to as channel knowledge map \cite{10287775,9373011,10764739}). Unlike existing channel knowledge maps \cite{10764739,10287775}, which aim to predict instantaneous channel parameters by mapping geographic locations to instantaneous channel realizations, the EM map here stores slowly varying statistical channel information. This approach is to reduce the overhead of channel reconstruction.

Specifically, analogous to the angle index graph constructed in \cite{10287775}, we assume the existence of an ideal EM map that contains the geometry and material data of the environment. The EM map represents the environment by a 3D EM model, denoted by $\omega$, which encodes the positions and EM properties of the BS, user terminals, and static reflectors or scatterers. The 3D EM model, $\omega$, can be reconstructed from sensors or preexisting geographic databases. Consequently, given antenna position control information, the statistical channel information in the EM map is expressed as
\begin{subequations}
	\begin{align}
		\Xi_{k,c,n}&=\{\bar{r}_{c,n}, \bar{r}_{k,c},\varsigma_{n,c}^{2}, \iota_{k,c}^{2}\} \\
		&= \mathrm{EMM}\big(\omega,\, \upsilon_k,\, p_n \big),
	\end{align}
\end{subequations}
for arbitrary antenna positions $p_n$, scatterers $c$, and user locations $\upsilon_k$, where $\mathrm{EMM}(\cdot)$ denotes the location-to-statistical-channel mapping function. The constructed EM map essentially serves as a database rooted in the location domain and establishes a mapping between user locations, antenna parameters, and statistical channel information, thereby guiding both antenna position optimization and mechanical beamforming design.

\subsubsection{Label Data Generation with Optimized Mechanical Beamforming} 
We generate multiple antenna-position samples
$
\{\mathbf{p}_j\}_{j=1}^{N_{\mathrm{s}}}
$
such that the antenna position space is uniformly covered, where the subscript $j$ is used to denote the index of the $j$-th training sample, and $N_{\mathrm{s}}$ is the total number of samples. Based on the EM map, we then generate the cluster-core channel $\tilde{\mathbf H}_{C}(\mathbf{p}_j)\in \mathbb{C}^{MN \times K}$ for every antenna-position vector $\mathbf{p}_j$. 
Moreover, for every $\mathbf{p}_j$, we draw $Z_{\mathrm{s}}$ statistical multipath channel realizations, denoted by $\{\tilde{\mathbf H}_{j,i}(\mathbf{p}_j)\}_{i=1}^{Z_{\mathrm{s}}}$ with $\tilde{\mathbf H}_{j,i}(\mathbf{p}_j)\in \mathbb{C}^{MN\times K}$, conditioned on cluster core and multipath statistics obtained in EM map. Specifically, for each cluster $c$, we draw per-realization sub-rays $\{ r_{l_c,n}, r_{k,c,l_c}\}$
around $\{\bar{r}_{c,n}, \bar{r}_{k,c}\}$ according to $\{\varsigma_{n,c}^{2}, \iota_{k,c}^{2}\}$. 

The resulting channel matrix samples can then be post-processed to extract key performance indicators. Specifically, given the set $\{\tilde{\mathbf H}_{j,i}(\mathbf p_j)\}_{i=1}^{Z_{\mathrm{s}}}$, we solve the fast-timescale ray-level problem (P2) to obtain the corresponding set of optimized  matrices $\{\mathbf G_j^*\}$.

Next, by substituting the obtained $\mathbf G_j^*$, $\mathbf p_j$, and the cluster-core channel $\tilde{\mathbf h}_{\mathrm{C},k}(\mathbf p_j)$ into the objective function of problem~(P3), we obtain the sum rate of all users for position $\mathbf p_j$ as $\bar{R}_j$.
Then, the tuple  
\(
\bigl(
\tilde{\mathbf H}_{j}(\mathbf{p}_j),
\bar R_j
\bigr)
\) is referred to as the $j$-th training sample, where $\tilde{\mathbf H}_{j}(\mathbf{p}_j)=\{\{\tilde{\mathbf H}_{j,i}(\mathbf{p}_j)\}_{i=1}^{Z_{\mathrm{s}}},\tilde{\mathbf H}_{C}(\mathbf{p}_j)\}$. 
We repeat the above process multiple times to generate the entire training data set.

\subsubsection{Slow-Fast Performance Surrogate Pre-Training}
The SFP surrogate is a data-driven function that takes statistical cluster-core and multipath channel 
as input and directly predicts the corresponding optimized user sum rate, which already accounts for the fast-timescale mechanical-beamforming optimization. In the following, the elements of the chosen deep neural network model for SFP surrogate training are described.

The proposed deep neural network architecture is designed as a Multi-Layer Perceptron network, which is well-established as a universal function approximator \cite{hornik1989multilayer}. The proposed network consists of $\bar{Q}$ layers, as illustrated in Fig. \ref{learning}. The network alternates between fully connected and nonlinear layers. The $q$-th layer in the network has a stack of $N_q$ neurons. For the non-linearity layers, Rectified Linear Units (ReLUs) are employed \cite{goodfellow2016deep}.
The input of the network is the real and imaginary parts of channel  $\tilde{\mathbf H}$, and the output of the network is the sum rate $\bar{R}$. This model learns how to map an input (i.e., sampled channel vector) to an output (i.e., predicted achievable rate).
All samples are normalized using dataset-level scaling \cite{goodfellow2016deep}, i.e.,
$\tilde{\mathbf H}^{\mathrm{norm}} = \frac{\tilde{\mathbf H}}{\Delta}$ with 
$\Delta = \max_{i,j}\lvert[\tilde {\mathbf H}]_{i,j}\rvert$.
This normalization choice preserves distance information encoded in the environment descriptors. Since modern implementations of deep learning models mainly use real-valued computations, each complex entry of the input data is split into real and imaginary values. 
For output normalization, let $R_{\max}=\max_{j\in\mathcal J}{\bar R_j}$ be the largest sum-rate value in the training set, where $\mathcal J=\{1,2,\cdots,N_{\mathrm{s}}\}$. Prior to training, each label is normalized as
$\frac{\bar R_j}{R_{\max}}, \forall j\in\{1,2,\cdots, N_{\mathrm{s}}\}$.

Let \(F_{\boldsymbol{\tau}}(\tilde{\mathbf H}(\mathbf{p})):\mathbb R^{MN}\!\to\!\mathbb R\) denote the proposed neural network with parameters $\boldsymbol{\tau}$. We use the entire training data set, $
\{
\tilde{\mathbf H}_{j}(\mathbf{p}_j),
\bar R_j
\bigr\}$,
to optimize the weights of the neural network $\boldsymbol{\tau}$. The cost function is defined as the mean squared error (MSE) between the label ${\bar R_j}$ and the neural network output. Specifically, the model is trained using a regression loss function. The training is guided through minimizing the following loss function
\begin{align}
	F_{\boldsymbol{\tau}}=
	\frac{1}{N_{\mathrm{s}}}
	\sum_{j=1}^{N_{\mathrm{s}}}
	\bigl(
	F_{\boldsymbol{\tau}}(\tilde{\mathbf H}_j(\mathbf{p}_j))
	-\bar R_j
	\bigr)^{2}.
	\label{eq:MSE}
\end{align}
After convergence, surrogate $F_{\boldsymbol{\tau}}$ accurately
predicts the expected sum-rate that is already maximized over $\mathbf G$, rendering the inner mechanical beamforming optimization loop unnecessary online.

\subsubsection{Fine-Tuning for Accurate Position Optimization}
Wireless networks exhibit inherent dynamics in both large-scale and small-scale statistics. In our two-timescale framework, cluster-core statistics vary slowly, whereas intra-cluster multipath-ray statistics fluctuate rapidly and are already handled by the fast-timescale mechanical-beamforming optimizer for $\mathbf G$ in problem (P2). Consequently, the offline-learned SFP surrogate may become mismatched at the slow timescale, leading to performance degradation when applied online. To mitigate this issue, we adopt a transfer-learning strategy \cite{9569500}, where the online deep neural network (DNN) is initialized by the offline SFP model and subsequently fine-tuned using a limited number of newly labeled samples from the current environment.

Specifically, we collect a small set of samples that capture the present channel conditions:
\begin{align}\label{ret}
	\bigl\{\mathbf H^{\mathrm{W}}_j(\mathbf p_j),\, \bar R^{\mathrm{W}}_j\bigr\}_{j=1}^{N_{\mathrm{w}}}, \qquad N_{\mathrm{w}}\ll N_{\mathrm{s}},
\end{align}
where $\mathbf H^{\mathrm{W}}_j(\mathbf p_j)$ denotes the collected channel samples for position $\mathbf p_j$, $\bar R^{\mathrm{W}}_j$ is the corresponding average sum rate, and $N_{\mathrm{w}}$ is the number of newly collected 
channel samples for fine-tuning, which is much smaller than the size of the offline training dataset $N_{\mathrm{s}}$.

Then, we refine the pre-trained SFP network via domain adaptation, where a subset of layers is frozen, and only the remaining layers are fine-tuned so that the DNN adapts to the current cluster-core channel coefficients. Mathematically,
\begin{align}
	F_{\bar{\boldsymbol{\tau}}}\!\big(\mathbf H^{\mathrm{W}}(\mathbf p)\big)
	= F^{\mathrm{FL}}_{\bar{\boldsymbol{\tau}}}\!\left(
	F^{\mathrm{TL}}_{\boldsymbol{\tau}^*}\!\big(\mathbf H^{\mathrm{W}}(\mathbf p)\big)\right),
	\label{eq:transfer_learning}
\end{align}
where $F^{\mathrm{TL}}_{\boldsymbol{\tau}^*}(\cdot)$ denotes the frozen layers with parameters ${\boldsymbol{\tau}}^{*}$ obtained in the pre-training stage, and $F^{\mathrm{FL}}_{\bar{\boldsymbol{\tau}}}(\cdot)$ denotes the fine-tuned layers with updated parameters $\bar{\boldsymbol{\tau}}$. Because only the small set $\{\mathbf H^{\mathrm{W}}_j(\mathbf p_j),\,R^{\mathrm{W}}_j\}$ is required, the online retraining time meets the stringent latency requirements of the flexible coupler system.

Replacing the objective function in problem (P3) by the learned SFP surrogate yields
\begin{subequations}\label{sfpp}
	\begin{align}
		\text{(P4)}\;&\max_{\mathbf p}\;
		F_{\bar{\boldsymbol{\tau}}}\!\big(\mathbf H^{\mathrm{W}}(\mathbf p)\big) \\[2pt]
		\text{s.t.}\;& 0\le p_n\le X,\ \forall n, \label{bt5}\\
		& |p_n-p_{n'}|\ge d_{\min},\ \forall n\neq n'. \label{bt6}
	\end{align}
\end{subequations}
where $F_{\bar{\boldsymbol{\tau}}}(\cdot)$ is a trained neural surrogate evaluated numerically. Based on problem (P4), the BS observes the current cluster-core CSI feature and updates $\mathbf p$ via projected gradient ascent \cite{hassani2017gradient}. By indexing the flexible coupler antennas along the rail in nondecreasing order of their positions (i.e., $p_1\le \cdots \le p_N$), the nonconvex constraints \eqref{bt6} can be simplified as the following linear constraints:
\begin{align}
	{p}_n- {p}_{n-1} \geq d_{\min},  \forall n =2, \cdots, N.
	\label{eq:linear_constraints}
\end{align}
Then, we define
\begin{align}
	\mathcal T\! \triangleq \! \Big\{\mathbf p\in\mathbb R^{N}: 0\!\leq\!  p_{1}, ~ p_{n+1}-p_{n}\!\ge \!d_{\min}\ (n\!=\!1,&\ldots, N-1),\nonumber
	&\\ p_{N}\le X \Big\},
\end{align}
as a nonempty and closed convex polyhedron. Note that $\mathcal T$ is guaranteed to be nonempty since $X \ge (N-1)d_{\min}$.
Next, for iteration $t$, we execute
\begin{subequations}
	\begin{align}
		\mathbf e^{(t)} &=
		\nabla_{\!\mathbf p}\,
		F_{\bar{\boldsymbol{\tau}}}\!\big(\mathbf H^{\mathrm{W}}(\mathbf p^{(t)})\big),
		\label{eq:proj_step1}\\
		\mathbf p^{(t+1)} &= \operatorname{Proj}_{\mathcal T}\!\big(
		\mathbf p^{(t)} + \eta\,\mathbf e^{(t)}\big),
		\label{eq:proj_step}
	\end{align}
\end{subequations}
and terminate when $\lVert\mathbf e^{(t)}\rVert<\varepsilon$, where $\varepsilon$ is a predefined tolerance and $\eta$ is the fixed step size of the projected-gradient update \cite{10130156}. Here, the Euclidean projection $\operatorname{Proj}_{\mathcal T}(\mathbf x)$ maps any vector $\mathbf x$ onto the closed convex set $\mathcal T$ by returning the unique point in $\mathcal T$ that minimizes the Euclidean distance to $\mathbf x$.

\begin{algorithm}[t]
	\caption{Digital Agent-Assisted Two-Timescale Optimization Algorithm for Flexible Coupler Array}
	\label{alg:LCDT}
	\begin{algorithmic}[1]
		\STATE \textbf{Inputs:} Number of samples $N_{\mathrm{s}}$; step size $\eta$; tolerance $\varepsilon$.
		\STATE \textbf{Outputs:} Antenna position vector $\mathbf{p}^{*}$ and mechanical beamformer $\mathbf G^*$.
		\STATE \textbf{Step 1: Slow timescale position optimization}
		\begin{itemize}
			\item \textbf{Antenna digital agent pre-training}
			\begin{enumerate}
				\item EM map construction.
				\item Generate optimized mechanical beamforming matrices $\{\mathbf G_{j}^{*}\}$ by running {\bf Step 2} and generate label data.
				\item Calculate the average sum rate $\bar{R}_j$ for $\mathbf{p}_j$.
				\item Repeat the above process multiple times to generate the entire training data set 
				$\{		
				\tilde{\mathbf H}_{j}(\mathbf{p}_j),
				\bar R_j\}_{j=1}^{N_{\mathrm{s}}}$.
				\item A neural network \(F_{\boldsymbol{\tau}}(\tilde{\mathbf H}(\mathbf{p}))\) is trained by minimizing \eqref{eq:MSE} to serve as a surrogate for SFP.
			\end{enumerate}
		\end{itemize}
		\begin{itemize}
			\item \textbf{Antenna digital agent online fine-tuning and control} :\\
			\begin{enumerate}
				\item Refresh statistical channel state information with the current environment.
				\item Collect retraining data from \eqref{ret}.
				\item Update the fine-tuned DNN parameters $\bar{\boldsymbol\tau}$ via \eqref{eq:transfer_learning}.
				\item \textbf{while} $\|\mathbf e^{(t)}\| >\varepsilon$ \textbf{do}\\
				Update $\mathbf e^{(t)}$ and $\mathbf{p}^{(t+1)}$ via \eqref{eq:proj_step1} and \eqref{eq:proj_step}, respectively.
				\item \textbf{end while}
				\item Set $\mathbf{p}^{*} = \mathbf{p}^{(t+1)}$.
			\end{enumerate}
		\end{itemize}
		\STATE \textbf{Step 2: Fast timescale mechanical beamforming with given $\mathbf{p}^*$}
		\begin{itemize}
				\item \textbf{while} no convergence of $\mathbf{S}$ \textbf{do}
			\begin{enumerate}
				\item Update $\mathbf{v}$ by solving problem \eqref{fff}.
				\item Update radiation pattern selection vector $\mathbf{s}_n$ by  \eqref{pp1}.
			\end{enumerate}
			\item \textbf{end while}
			\item Construct the mechanical beamformer $\mathbf{G}^*$ by \eqref{eq:23}.
		\end{itemize}
	\end{algorithmic}
\end{algorithm}

The overall digital agent-assisted two-timescale optimization algorithm for flexible coupler array is detailed in Algorithm 1. The algorithm guarantees convergence to local optimal points since the objective value is non-decreasing in each iteration. In Algorithm 1, the computational complexity of antenna position optimization is $\mathcal{O}(2K+2LMN^3+2KLN)$ where $L = \max\{L_1, L_2, \ldots, L_K\}$
\cite{9569500}. 

\section{Simulation Results}
\begin{table}[t]
	\centering
	\caption{Simulation Parameters.}
	\begin{tabular}{|m{1.4cm}|m{5.0cm}|m{0.8cm}|}
		\hline
		\textbf{Parameter} & \textbf{Description} & \textbf{Value} \\
		\hline
		$\eta$ & Learning rate for training the DNN & 0.01 \\
		\hline
		$S_B$ & Number of training iterations for pre-training & 1000 \\
		\hline
		$S_T$ & Number of training iterations for fine tuning & 100 \\
		\hline
		$B$ & Batch size for split training and testing sets & 32 \\
		\hline
		$\epsilon$ & Smoothing term to prevent division by zero for Adam & $10^{-8}$ \\
		\hline
		$N_{\mathrm{s}}$ & Number of training samples used for pre-training & $10^5$ \\
		\hline
		$N_{\mathrm{W}}$ & Number of training samples used for retraining in a mismatch scenario & $10^3$ \\
		\hline
	\end{tabular}
	\label{tab:hyperparameters}
\end{table}
In the simulation, the users are randomly distributed within a two-dimensional rectangular region of size $S_x=8$ m by $S_y=15$ m. The total number of flexible coupler antennas is \(N=16\). The carrier frequency is 28 GHz. We set $U=14$ and $M=360$.
In this scenario, all $K=10$ users operate under a scattering-cluster channel with $C_k=3$ clusters, each consisting of $L_{c,k}=3$ multipaths. The nominal distances of these scattering clusters, $\bar{r}_{c,n}$ and $\bar{r}_{k,c}$, are randomly drawn from the range $[2, 4]$ m, while the distance spreads are set to $\varsigma_{n,c}=0.4$ m and $\iota_{k,c}=0.4$ m.
In the proposed digital agent algorithm, the
DNN is composed of four dense layers, with
the number of neurons as 500, 250, 100, and 50, respectively. The EM map is constructed via intelligent ray tracing \cite{1040665}, where physical paths are grouped in the joint angle and delay domain. For each cluster, we compute the mean delay and the delay variance. These statistics determine the cluster core distances and the corresponding distance variance parameters for each user and antenna position.

We consider the traditional two-timescale optimization structure as the baseline scheme. At the fast timescale, the BS first estimates the statistical CSI and generates multipath channel samples to evaluate the ergodic sum rate, and then obtains the mechanical beamforming matrix by solving problem (P2). At the slow  timescale, the BS first computes the cluster-core CSI and then iteratively updates the antenna-position vector $\mathbf{p}^{(t)}$ using the conditional gradient algorithm \cite{shao20246d}. At each outer iteration of the position update, it calls the fast mechanical beamforming subproblem (P2) as the inner loop to obtain $\mathbf{G}^{(t)}$, and the process continues until $\mathbf{p}^{(t)}$ converges. The complexity of updating the slow-timescale position vector \(\mathbf{p}\) is
$
\mathcal{O}\bigl(I\,(T_\mathrm{H}\,J(\sqrt{2UN}\,\log \tfrac{1}{\xi} ( (UN)^3 + Z_{\mathrm{s}}K^2UN + Z_{\mathrm{s}}K^3 )) + NK^2)\bigr)$ \cite{shao20246d},
where \(I\) denotes the number of iterations in the position optimization, \(J\) denotes the number of iterations in the mechanical beamforming, and $T_H$ denotes the number of fast channel samples used per slow-timescale position update.  
\begin{figure}[!t]
	\centering
	\includegraphics[width=0.939\linewidth]{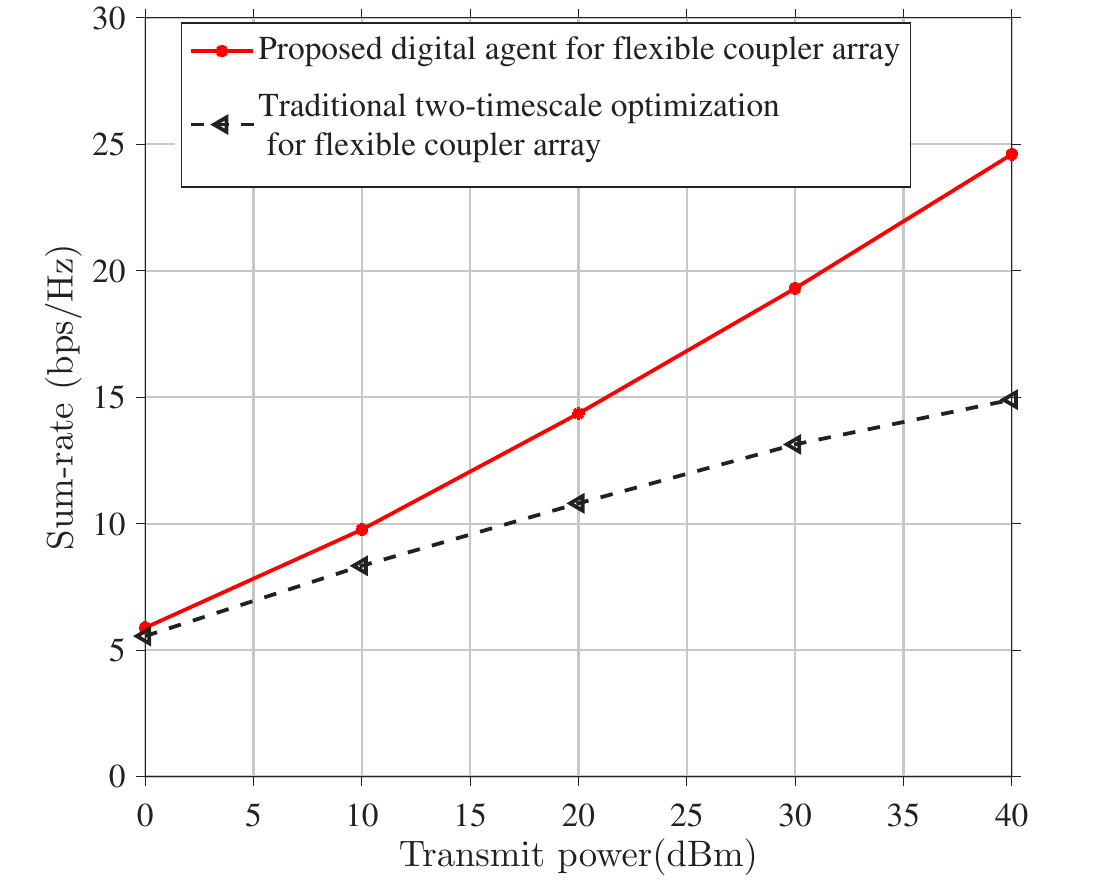}
	\caption{Comparison of the digital agent and traditional two-timescale optimization algorithms.}
	\label{fig:sumrate}
		\vspace{-0.59cm}
\end{figure} 

Fig.~\ref{fig:sumrate} compares the proposed digital-agent algorithm with the traditional two-timescale optimization for flexible coupler systems under a deliberately mismatched cluster-core distribution. In this scenario, the nominal distance range of the scattering clusters for each user is enlarged to $[3,5]$ m, while $\{C_k\}$ and $\{L_{c,k}\}$ remain unchanged. The traditional baseline fails to adapt to channel variations, which results in a sharp decline in the average sum rate. In contrast, the digital agent adapts well to channel information and simultaneously reduces the online complexity of the traditional two-timescale methods to $\mathcal{O}(2K+2LMN^3+2KLN)$. For example, the proposed digital agent algorithm achieves $95\%$ of the final average sum rate within approximately $32.4$~s, 
whereas the baseline requires approximately $10.9$~min, 
since each slow-timescale update requires rerunning the inner fast-timescale loop.
 When the channel statistics drift, the agent fine-tunes only the last two network layers with a small number of labeled channel realizations, which preserves adaptability without incurring large online cost. These results indicate that the proposed digital agent ensures reliable performance by solving the optimization within the digital agent rather than relying entirely on the physical environment.
\begin{figure}[!t]
	\centering
	\includegraphics[width=0.91\linewidth]{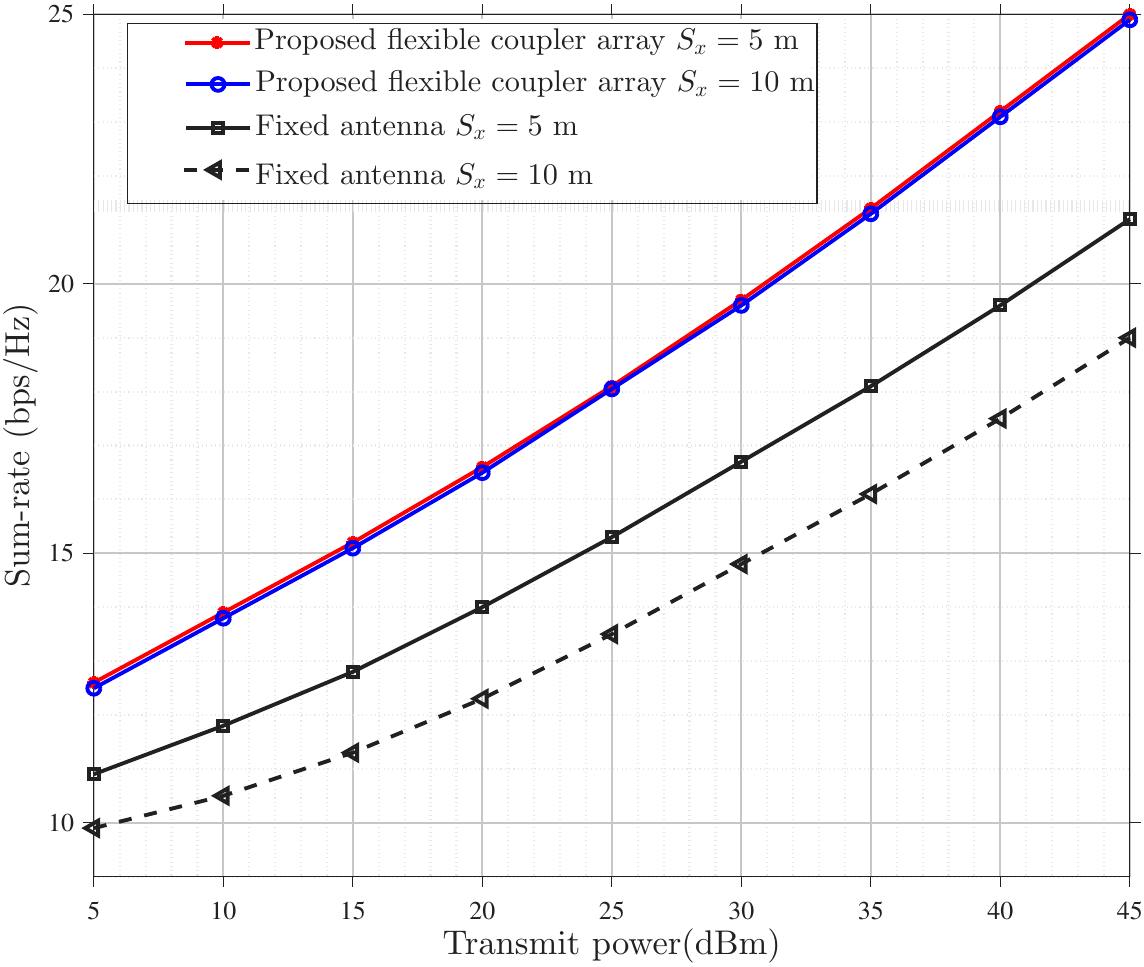}
	\caption{Impact of the reserved area’s horizontal length $S_x$ on flexible coupler system performance.}
	\label{sx}
		\vspace{-0.59cm}
\end{figure} 

Fig.~\ref{sx} illustrates the average sum rate as a function of transmit power for the proposed flexible coupler array and the conventional fixed antenna with fixed position, rotation, and radiation pattern, evaluated for deployment areas with horizontal lengths of $S_x = 5$ m and $S_x = 10$ m.
As observed, the proposed flexible coupler array consistently achieves
a higher average rate compared to the conventional system, which
demonstrates its ability to provide enhanced communication
performance. In addition, the average rate of the proposed flexible coupler array remains
nearly identical for both deployment areas, while the conventional
system experiences a significant degradation when transitioning
from $S_x= 5$ m to $S_x= 10$ m. This advantage is attributed to the flexibility of the flexible coupler antenna to adjust its  position along the slide rail to be closer to the user, effectively mitigating path loss and maintaining strong  communication links. This behavior highlights that the flexible coupler performance is less sensitive to variations in deployment area, which confirms that dynamic positioning effectively mitigates the adverse effects of increased propagation distance. In contrast, the conventional system suffers significant performance degradation due to the fixed position of its antenna, which amplifies the path loss in wireless propagation as the horizontal length $S_x$ increases. 
\begin{figure}[!t]
	\centering
	\includegraphics[width=0.91\linewidth]{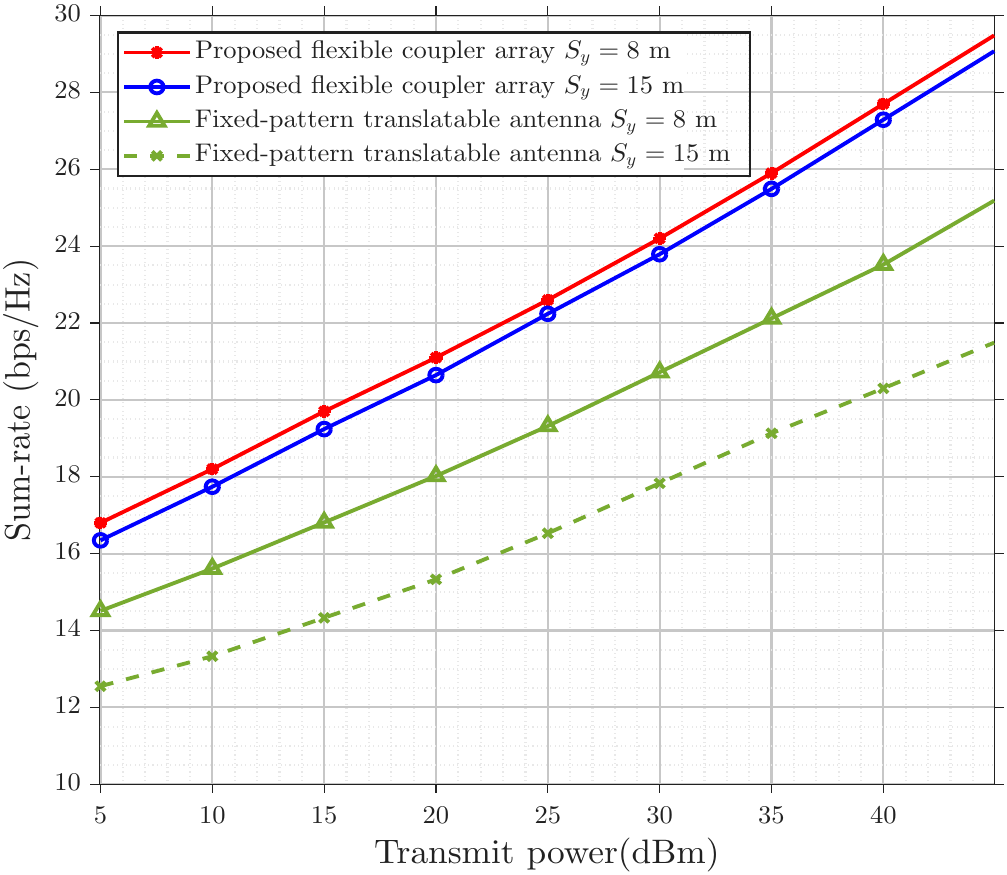}
	\caption{Impact of the reserved area’s vertical length $S_y$ on flexible coupler system performance.}
	\label{sy}
\end{figure} 

Fig. \ref{sy} plots the average sum-rate versus transmit power for the proposed flexible coupler system and fixed-pattern translatable antenna system (i.e., an antenna with a fixed rotation and pattern that can only slide along a rail), evaluated for deployment areas of vertical length $S_y = 8$ m and $S_y = 15$ m. It can be observed that extending the reserved area increases the propagation distance, thereby degrading the throughput of the fixed-pattern translatable antenna. In contrast, the two curves corresponding to the proposed flexible coupler array remain close, with only minor differences attributable to the additional path loss introduced by the taller service region. These results verify that the combination of sliding mobility and radiation pattern reconfiguration of couplers effectively mitigates the geometric penalty of a larger vertical coverage area, thereby sustaining nearly optimal performance across the entire coverage range. It is worth noting that the robustness of the proposed flexible coupler array is achieved using a single slide rail, as each antenna element can both slide and reshape its radiation pattern. In comparison, the fixed-pattern translatable antenna design typically requires multiple parallel rails to achieve comparable rates, which substantially increases hardware complexity and deployment cost.

\begin{figure}[!t]
	\centering
	\includegraphics[width=0.91\linewidth]{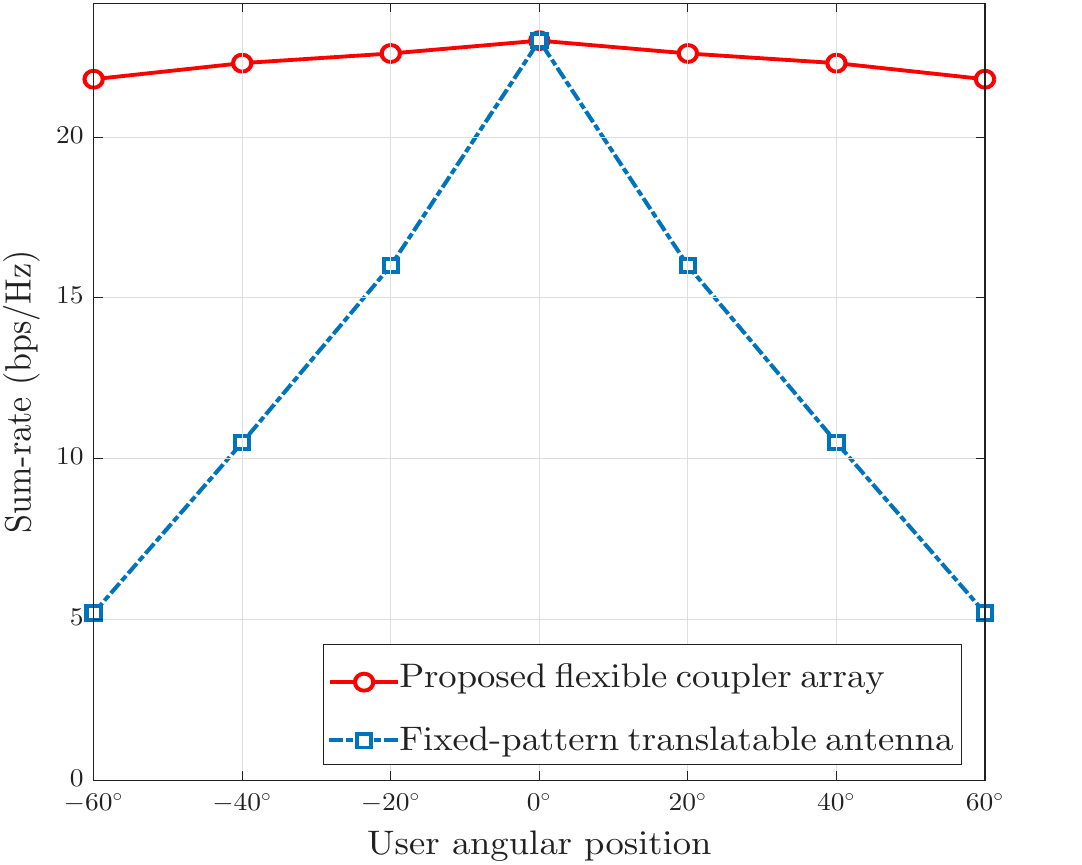}
	\caption{The sum-rate of flexible coupler system versus the angle of the user.}
	\label{angle}
		\vspace{-0.59cm}
\end{figure} 

Fig. \ref{angle} illustrates the sum-rate variation with respect to the user’s angular position relative to the flexible coupler array, where $K=1$ and $N=1$. The proposed flexible coupler array maintains robust performance across a wide range of user angles by dynamically adjusting the pattern orientations as the user position deviates from the optimal direction.
In contrast, the fixed-pattern translatable antenna achieves satisfactory performance only when the user is positioned perpendicularly to the array, where the antenna gain is maximized. These results underscore the robustness and versatility of the proposed mechanical beamforming approach, demonstrating that dynamic pattern reconfiguration not only compensates for variations in user location but also substantially enhances the overall system performance and reliability across diverse deployment scenarios.

\begin{figure}[!t]
	\centering
	\includegraphics[width=0.91\linewidth]{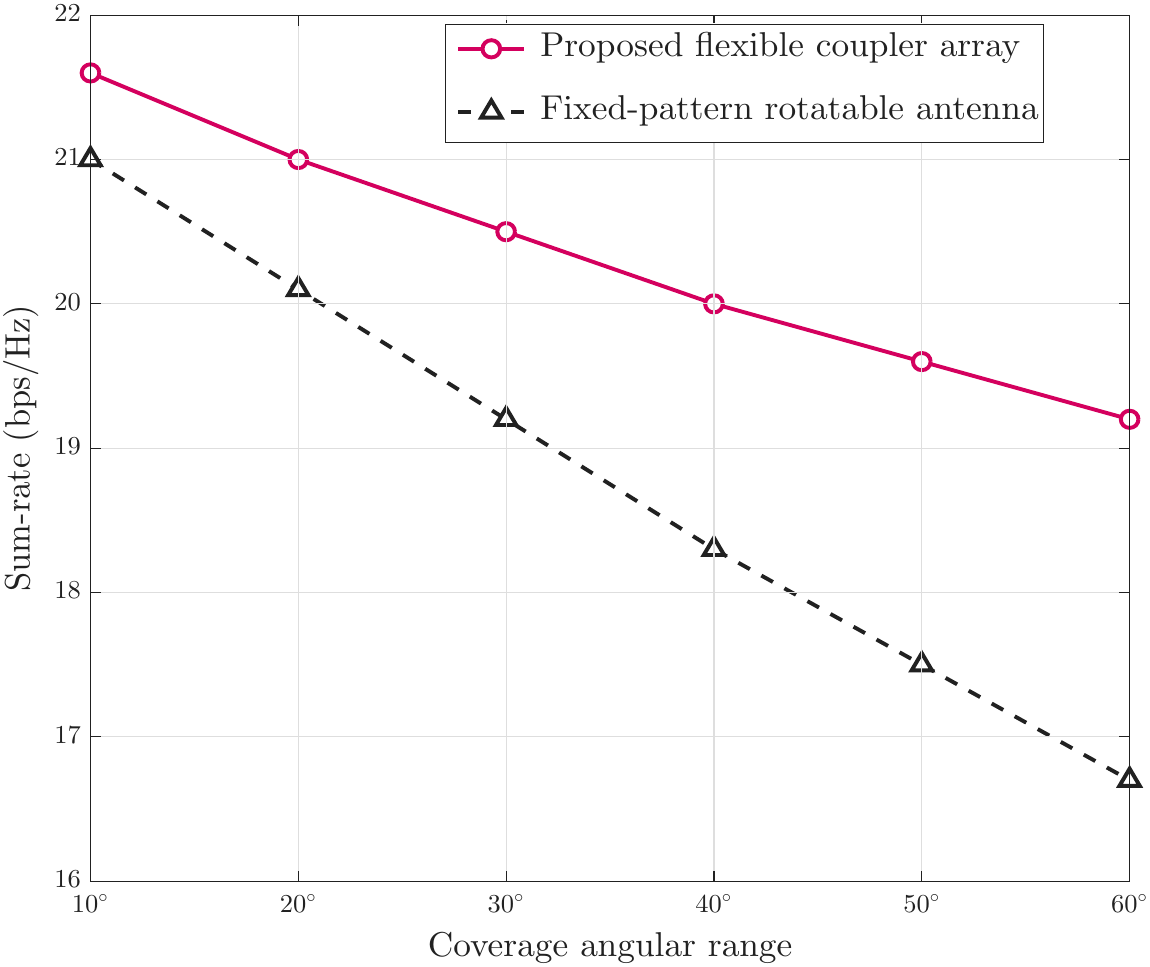}
	\caption{The sum-rate of flexible coupler system versus the coverage angular range.}
	\label{emr}
	\vspace{-0.59cm}
\end{figure}

Fig. \ref{emr} presents the average sum rate versus the total user coverage angle for two antenna configurations, namely proposed flexible coupler array with fixed active antenna positions and the rotatable antenna with a fixed radiation pattern (i.e., the active antenna remains at a fixed position with a fixed pattern, but it can rotate around its own axis). We can see that when the coverage angular range is small, both schemes achieve similar performance because their main lobes can remain aligned with the users. As the angle increases (i.e., the users are more widely dispersed), the fixed-pattern rotatable antenna experiences rapid performance degradation due to misalignment. By contrast, the proposed flexible coupler array exhibits a much slower performance decline, since coupler movement keeps the array oriented toward the prevailing user sector, while pattern reconfiguration compensates for residual offsets and angular dispersion among users. These results indicate that flexible coupler array is essential for sustaining high sum-rate performance over wide angular coverage.

\section{Conclusions}
We have proposed a flexible coupler array that combines the
small-range relocation of passive couplers with large-scale antenna translation. The array introduces an additional radiation DoF with relocating couplers and reduces large-scale path loss by enabling near-user placement. We have then designed a two-timescale optimization framework that maximizes the sum rate by jointly optimizing antenna position at the slow timescale and mechanical beamforming at the fast timescale. To solve this problem, we have proposed a digital-agent-aided optimization algorithm to reduce online computational complexity. The proposed coupler position reconfiguration provides a cost-effective way to enhance communication performance without moving active antennas, thereby improving antenna adaptability with a compact structure and low hardware overhead. For our future work, we will investigate channel estimation for the flexible coupler array, since each new configuration of passive coupler positions results in a different channel as well as varied coupling with the active antenna element.

\bibliographystyle{IEEEtran}
\bibliography{fabs}

@ARTICLE{9724202,
	author={Shao, Xiaodan and You, Changsheng and Ma, Wenyan and Chen, Xiaoming and Zhang, Rui},
	journal={IEEE J. Sel. Areas Commun.}, 
	title={Target Sensing With Intelligent Reflecting Surface: Architecture and Performance}, 
	year={2022},
	volume={40},
	number={7},
	pages={2070-2084},
	month={Jul.},}

@ARTICLE{10443321,
	author={Shao, Xiaodan and Zhang, Rui},
	journal={IEEE Trans. Wireless Commun.}, 
	title={Target-Mounted Intelligent Reflecting Surface for Secure Wireless Sensing}, 
	year={2024},
	volume={23},
	number={8},
	pages={9745-9758},
	month={Aug.},}

@ARTICLE{8937497,
	author={Shao, Xiaodan and Chen, Xiaoming and Jia, Rundong},
	journal={IEEE Trans. Signal Process.}, 
	title={A Dimension Reduction-Based Joint Activity Detection and Channel Estimation Algorithm for Massive Access}, 
	year={2020},
	volume={68},
	number={},
	pages={420-435},
	month={Dec.},}

@ARTICLE{10643661,
	author={Li, Mushu and Gao, Jie and Zhou, Conghao and Zhao, Lian and Shen, Xuemin},
	journal={IEEE Internet Things J.}, 
	title={Digital-Twin-Empowered Resource Allocation for On-Demand Collaborative Sensing}, 
	year={2024},
	volume={11},
	number={23},
	pages={37942-37958},
	month={Dec.},}

@ARTICLE{11025157,
	author={Chen, Nan and Ye, Qiang and Rashid, Mamunur and Zheng, Yang},
	journal={IEEE Internet Things Mag.}, 
	title={Realizing Sustainable and Adaptive Smart Cities With AI-Powered Digital Twin}, 
	year={2025},
	volume={8},
	number={5},
	pages={37-44},
	month={Sep.},}

@book{cvx,
		title={Convex optimization algorithms},
		author={Bertsekas, Dimitri},
		year={2015},
		publisher={Athena Scientific}
	}

@ARTICLE{753729,
	author={Fleury, B.H. and Tschudin, M. and Heddergott, R. and Dahlhaus, D. and Ingeman Pedersen, K.},
	journal={IEEE J. Sel. Areas Commun.}, 
	title={Channel parameter estimation in mobile radio environments using the {SAGE} algorithm}, 
	year={1999},
	volume={17},
	number={3},
	pages={434-450},
	ISSN={1558-0008},
	month={Mar.},}

@ARTICLE{10287775,
	author={Wu, Di and Zeng, Yong and Jin, Shi and Zhang, Rui},
	journal={IEEE Trans. Wireless Commun.}, 
	title={Environment-Aware Hybrid Beamforming by Leveraging Channel Knowledge Map}, 
	year={2024},
	volume={23},
	number={5},
	pages={4990-5005},
	month={May},}

@INPROCEEDINGS{1040665,
	author={Rautiainen, T. and Wolfle, G. and Hoppe, R.},
	booktitle={Proc. IEEE Veh. Technol. Conf.}, 
	title={Verifying path loss and delay spread predictions of a {3D} ray tracing propagation model in urban environment}, 
	year={2002},
	volume={4},
	number={},
	pages={2470-2474},
	month={Sept.},}

@ARTICLE{6691942,
	author={Jalali Mazlouman, Shahrzad and Mahanfar, Alireza and Soleimani, Maryam and Chan, Henry and Menon, Carlo and Vaughan, Rodney G.},
	journal={IEEE Trans. Antennas Propag.}, 
	title={Pattern Reconfiguration by Rotating Parasitic Structure Using Electro-Active Polymer {(EAP)} Actuator}, 
	year={2014},
	volume={62},
	number={3},
	pages={1046-1055},
	month={Mar.},}

@article{hornik1989multilayer,
	title={Multilayer feedforward networks are universal approximators},
	author={Hornik, Kurt and Stinchcombe, Maxwell and White, Halbert},
	journal={Neural networks},
	volume={2},
	number={5},
	pages={359--366},
	year={1989},
	publisher={Elsevier}
}

@ARTICLE{10130156,
	author={Chen, Jung-Chieh and Lin, Yu-Cheng},
	journal={IEEE Access}, 
	title={A Projected Gradient Descent Algorithm for Designing Low-Resolution Finite-Alphabet Equalizers in All-Digital Massive {MU-MIMO} Communication Systems}, 
	year={2023},
	volume={11},
	number={},
	pages={50744-50751},
	month={May},}

@book{goodfellow2016deep,
	title={Deep learning},
	author={Goodfellow, Ian and Bengio, Yoshua and Courville, Aaron and Bengio, Yoshua},
	volume={1},
	number={2},
	year={2016},
	publisher={MIT press Cambridge}
}

@article{wen,
	title={Aerial {6D} Movable Antenna-Enabled Cell-Free Networks},
	author={Wei, Hao and Wang, Wen and Ni, Wanli and Zhang, Cheng and Huang, Yongming},
	journal={IEEE Trans. Veh. Technol. },
	volume={},
	number={},
	pages={},
	note={early access},
	year={2025}
}

@ARTICLE{10945745,
	author={Shao, Xiaodan and Zhang, Rui},
	journal={IEEE Commun. Mag.}, 
	title={{6DMA} Enhanced Wireless Network with Flexible Antenna Position and Rotation: Opportunities and Challenges}, 
	year={2025},
	volume={63},
	number={4},
	pages={121-128},
	month={Apr.},}

@ARTICLE{passive6DMA,
	author={Wang, Haozhe and Shao, Xiaodan and Zheng, Beixiong and Shi, Xiaoming and Zhang, Rui},
	journal={IEEE Wireless Commun. Lett.}, 
	title={Passive Six-Dimensional Movable Antenna {(6DMA)}-Assisted Multiuser Communication}, 
	year={2025},
	volume={14},
	number={4},
	pages={1014-1018},
	month={Apr.},}

@article{akritas1996various,
		title={Various proofs of Sylvester's (determinant) identity},
		author={Akritas, Alkiviadis G and Akritas, Evgenia K and Malaschonok, Genadii I},
		journal={Mathematics and Computers in Simulation},
		volume={42},
		number={4-6},
		pages={585--593},
		year={1996},
		publisher={Elsevier}
	}

@article{near,
	author={Shao, Xiaodan and others},
	journal={IEEE Trans. Wireless Commun.}, 
	title={Hybrid Near-Far Field {6D} Movable Antenna Design Exploiting Directional Sparsity and Deep Learning}, 
	volume={},
	number={},
	pages={},
	year={2025},
	note    = {Early access},}

@article{IPA,
	title   = {Polarforming Antenna Enhanced Sensing and Communication: Modeling and Optimization},
	author  = {Shao, Xiaodan and Zhang, Rui and Zhou, Haibo and Jiang, Qijun and Zhou, Conghao and Zhuang, Weihua and Shen, Xuemin},
	journal = {IEEE J. Sel. Areas Commun.},
		note    = {Early access},
	year    = {2025},
}

@ARTICLE{10989638,
	author={Lu, Haiquan and Yu, Zhi and Zeng, Yong and Ma, Shaodan and Jin, Shi and Zhang, Rui},
	journal={IEEE Trans. Wireless Commun.}, 
	title={Wireless Communication With Flexible Reflector: Joint Placement and Rotation Optimization for Coverage Enhancement}, 
	year={2025},
	volume={24},
	number={10},
	pages={8252-8266},
	month={Oct.},}

@ARTICLE{11214475,
	author={Qiao, Luteng and Zhou, Yong and Shi, Yuanming and Wu, Sheng and Luan, Tom H.},
	journal={IEEE Trans. Veh. Technol.}, 
	title={{UAV}-Assisted Mobile Edge Computing With Digital Twin}, 
	year={2025},
	volume={74},
	number={11},
	pages={17985-17998},
	month={Nov.},}

@article{moon2022deep,
	title={Deep neural network for beam and blockage prediction in {3GPP}-based indoor hotspot environments},
	author={Moon, Sangmi and Kim, Hyeonsung and You, Young-Hwan and Kim, Cheol Hong and Hwang, Intae},
	journal={Wirel. Pers. Commun.},
	volume={124},
	number={4},
	pages={3287--3306},
	year={2022},
	publisher={Springer}
}

@inproceedings{lutzmayr2022wireless,
	title={Wireless Communication Technologies in Smart Factories},
	author={Lutzmayr, Dieter and Pauritsch, Manfred},
	booktitle={Inter. Conf. Syst. Integr. Intell.},
	pages={449--457},
	year={2022},
	organization={Springer}
}

@ARTICLE{8468019,
	author={Domínguez-Bolaño, Tomás and Rodríguez-Piñeiro, José and García-Naya, José A. and Yin, Xuefeng and Castedo, Luis},
	journal={IEEE Access}, 
	title={Measurement-Based Characterization of Train-to-Infrastructure 2.6 {GHz} Propagation Channel in a Modern Subway Station}, 
	year={2018},
	volume={6},
	number={},
	pages={52814-52830},
	month={Sept.},}

@article{pi20246d,
	title={{6D} Movable Antenna Enhanced Multi-Access Point Coordination via Position and Orientation Optimization},
	author={Pi, Xiangyu and Zhu, Lipeng and Mao, Haobin and others},
	journal={arXiv preprint arXiv:2412.10736},
	year={2024}
}

@ARTICLE{10436574,
	author={Xu, Hao and Wong, Kai-Kit and New, Wee Kiat and Tong, Kin-Fai and Zhang, Yangyang and Chae, Chan-Byoung},
	journal={IEEE Trans. Wireless Commun.}, 
	title={Revisiting Outage Probability Analysis for Two-User Fluid Antenna Multiple Access System}, 
	year={2024},
	volume={23},
	number={8},
	pages={9534-9548},
	month={Aug.},}

@ARTICLE{10638530,
	author={Zheng, Jinkai and Luan, Tom H. and Zhang, Yao and Li, Guanjie and Su, Zhou and Wu, Wen},
	journal={IEEE Wireless Commun.}, 
	title={Digital Twin in {6G}: Embracing Comprehensive Network Intelligence}, 
	year={2024},
	volume={31},
	number={6},
	pages={94-101},
	month={Dec.},}

@ARTICLE{proc,
	author={Wu, Qingqing and others},
	journal={Proc. IEEE}, 
	title={Intelligent Surfaces Empowered Wireless Network: Recent Advances and the Road to {6G}}, 
	year={2024},
	volume={112},
	number={7},
	pages={724-763},
	ISSN={1558-2256},
	month={Jul.},}

@ARTICLE{exl,
	author={Wang, Zhe and Zhang, Jiayi and Du, Hongyang and others},
	journal={IEEE Commun. Surv. Tutorials.},
	title={A Tutorial on Extremely Large-Scale {MIMO} for {6G}: {Fundamentals}, Signal Processing, and Applications},
	year={2024},
	volume={26},
	number={3},
	pages={1560-1605},
	month={thirdquart.},}

@ARTICLE{Larsson2014Massive,
		author={E.~G.~{Larsson} and O.~{Edfors} and F.~{Tufvesson} and T.~L.~{Marzetta}},
		journal={IEEE Commun. Mag.}, 
		title={Massive {MIMO} for next generation wireless systems}, 
		year={2014},
		volume={52},
		number={2},
		pages={186-195},
		month={Feb.}}

@ARTICLE{8122055,
	author={Li, Xingjian and Fang, Jun and Li, Hongbin and Wang, Pu},
	journal={IEEE Trans. Wireless Commun.}, 
	title={Millimeter Wave Channel Estimation via Exploiting Joint Sparse and Low-Rank Structures}, 
	year={2018},
	volume={17},
	number={2},
	pages={1123-1133},
	month={Feb.},}

@ARTICLE{9373011,
		author={Zeng, Yong and Xu, Xiaoli},
		journal={IEEE Wirel. Commun.}, 
		title={Toward Environment-Aware {6G} Communications via Channel Knowledge Map}, 
		year={2021},
		volume={28},
		number={3},
		pages={84-91},
		month={Jun.},}

@ARTICLE{9198125,
	author={Zhao, Ming-Min and Wu, Qingqing and Zhao, Min-Jian and Zhang, Rui},
	journal={IEEE Trans. Wirel. Commun.}, 
	title={Intelligent Reflecting Surface Enhanced Wireless Networks: Two-Timescale Beamforming Optimization}, 
	year={2021},
	volume={20},
	number={1},
	pages={2-17},
	month={Jan.},}

@ARTICLE{10764739,
	author={Wang, Xiucheng and Tao, Keda and Cheng, Nan and Yin, Zhisheng and Li, Zan and Zhang, Yuan and Shen, Xuemin},
	journal={IEEE Trans. Cogn. Commun. Netw.}, 
	title={RadioDiff: An Effective Generative Diffusion Model for Sampling-Free Dynamic Radio Map Construction}, 
	year={2025},
	volume={11},
	number={2},
	pages={738-750},
	month={Apr.},}

@INPROCEEDINGS{9569500,
	author={Khan, Saud and Durrani, Salman and Zhou, Xiangyun},
	booktitle={IEEE Int. Symp. Pers. Indoor Mob. Radio Commun. (PIMRC)}, 
	title={Transfer Learning Based Detection for Intelligent Reflecting Surface Aided Communications}, 
	year={2021},
	volume={},
	number={},
	pages={555-560},
	month={Sept.},}

@article{hassani2017gradient,
	title={Gradient methods for submodular maximization},
	author={Hassani, Hamed and Soltanolkotabi, Mahdi and Karbasi, Amin},
	journal={Advances in Neural Information Processing Systems},
	volume={30},
	year={2017}
}

@ARTICLE{10623528,
		author={Cheng, Nan and Wang, Xiucheng and Li, Zan and Yin, Zhisheng and Luan, Tom H. and Shen, Xuemin},
		journal={IEEE Netw.}, 
		title={Toward Enhanced Reinforcement Learning-Based Resource Management via Digital Twin: Opportunities, Applications, and Challenges}, 
		year={2025},
		volume={39},
		number={1},
		pages={189-196},
		month={Jan.},}

@article{jiang2025statistical,
	title={Statistical channel based low-complexity rotation and position optimization for {6D} movable antennas enabled wireless communication},
	author={Jiang, Qijun and Shao, Xiaodan and Zhang, Rui},
	journal={arXiv preprint arXiv:2504.20618},
	year={2025}
}

@article{zeng2025robust,
	title={Robust resource allocation for pinching-antenna systems under imperfect {CSI}},
	author={Zeng, Ming and Wang, Xianbin and Liu, Yuanwei and Ding, Zhiguo and Karagiannidis, George K and Poor, H Vincent},
	journal={arXiv preprint arXiv:2507.12582},
	year={2025}
}

@ARTICLE{MA,
  author={Zhu, Lipeng and Ma, Wenyan and Xiao, Zhenyu and Zhang, Rui},
journal={IEEE Trans. Wireless Commun.}, 
title={Performance Analysis and Optimization for Movable Antenna Aided Wideband Communications}, 
year={2024},
volume={23},
number={12},
pages={18653-18668},
month={Dec.},}

@article{shao2025tutorial,
author={Shao, Xiaodan and Mei, Weidong and You, Changsheng and others},
journal={IEEE Commun. Surv. Tutor.}, 
title={A Tutorial on Six-Dimensional Movable Antenna for {6G} Networks: Synergizing Positionable and Rotatable Antennas},
note={early access}, 
year={2025},
volume={},
number={},
pages={},
month={},}

@ARTICLE{Grieves2014,
	author={Grieves, M.},
	title={Digital twin: Manufacturing excellence through virtual factory replication},
	journal={White Paper},
	volume={1},
	number={2014},
	pages={1-7},
	year={2014}
}

@ARTICLE{10886938,
	author={Jiang, Shuaifeng and Qu, Qi and Pan, Xiaqing and Agrawal, Abhishek K. and Newcombe, Richard and Alkhateeb, Ahmed},
	journal={IEEE Open J. Commun. Soc.}, 
	title={Learnable Wireless Digital Twins: Reconstructing Electromagnetic Field With Neural Representations}, 
	year={2025},
	volume={6},
	number={},
	pages={1568-1590},
	month={},}

@ARTICLE{9906937,
	author={Zhou, Conghao and Gao, Jie and Li, Mushu and Shen, Xuemin and Zhuang, Weihua},
	journal={J. Commun. Inf. Netw.}, 
	title={Digital Twin-Empowered Network Planning for Multi-Tier Computing}, 
	year={2022},
	volume={7},
	number={3},
	pages={221-238},
	month={Sept.},}

@article{shao20246d,
	author={Shao, Xiaodan and Jiang, Qijun and Zhang, Rui},
	journal={IEEE Trans. Wireless Commun.}, 
	title={{6D} movable antenna based on user distribution: Modeling and optimization}, 
	year={2025},
	volume={24},
	number={1},
	pages={355-370},
	ISSN={1558-2248},
	month={Jan.},}

@ARTICLE{6dmasensing,
	author={Shao, Xiaodan and Zhang, Rui and Schober, Robert},
	journal={IEEE Wireless Commun. Lett.}, 
	title={Exploiting Six-Dimensional Movable Antenna for Wireless Sensing}, 
	year={2025},
	volume={14},
	number={2},
	pages={265-269},
	month={Feb.},}

@INPROCEEDINGS{gaohierarchical,
		author={Gao, Junjie and Yan, Bin and Ma, Rui and Wang, Zheng and Huang, Yongming},
		booktitle={IEEE Int. Conf. Commun. Technol.}, 
		title={Hierarchical Alternating Optimization for Enhanced {6DMA}-Assisted Over-the-Air Computation}, 
		year={2025},
		volume={},
		number={},
		pages={},
		month={Oct.},
	}

@article{yan2025six,
	title={Six-Dimensional Movable Antenna Enabled Wideband {THz} Communications},
	author={Yan, Wencai and Hao, Wanming and Fan, Yajun and Guo, Yabo and Wu, Qingqing and Li, Xingwang},
	journal={arXiv preprint arXiv:2510.25088},
	year={2025}
}

@article{wang20256d,
	title={{6D} Movable Holographic Surface Assisted Integrated Data and Energy Transfer: A Sensing Enhanced Approach},
	author={Wang, Zhonglun and Zhao, Yizhe and Hu, Gangming and Zheng, Yali and Yang, Kun},
	journal={arXiv preprint arXiv:2510.21137},
	year={2025}
}

@article{Shao2026FC,
	title={Coupler Position Optimization and Channel Estimation for Flexible Coupler Aided Multiuser Communication},
	author={Xiaodan Shao and Chuangye Shan and Weihua Zhuang and Xuemin Shen},
	journal={arXiv preprint arXiv:2602.11319},
	year={2026},
	month={Feb.},
}

@article{shen20256d,
	title={{6D} Movable Metasurface ({6DMM}) in Downlink {NOMA} Transmissions},
	author={Shen, Li-Hsiang},
	journal={arXiv preprint arXiv:2510.17502},
	year={2025}
}

@article{liu2024uav,
	title={{UAV}-enabled passive {6D} movable antennas: Joint deployment and beamforming optimization},
	author={Liu, Changhao and Mei, Weidong and Wang, Peilan and Meng, Yinuo and Ning, Boyu and Chen, Zhi},
	journal={arXiv preprint arXiv:2412.11150},
	year={2024}
}

@ARTICLE{6DMA_JSTSP,
	author={Shao, Xiaodan and Zhang, Rui and Jiang, Qijun and Park, Jihong and Quek, Tony Q. S. and Schober, Robert},
	journal={IEEE J. Sel. Topics Signal Process.}, 
	title={Distributed Channel Estimation and Optimization for {6D} Movable Antenna: Unveiling Directional Sparsity}, 
	year={2025},
	volume={19},
	number={2},
	pages={349-365},
	month={Mar.},}

@ARTICLE{free6DMA,
	author={Shi, Xiaoming and Shao, Xiaodan and Zheng, Beixiong and Zhang, Rui},
	journal={IEEE Wireless Commun. Lett.}, 
	title={{6DMA}-Aided Cell-Free Massive {MIMO} Communication}, 
	year={2025},
	volume={14},
	number={5},
	pages={1361-1365},
	month={May},}

@ARTICLE{11314850,
	author={Cheng, Nan and others},
	journal={IEEE Trans. Wireless Commun.}, 
	title={Channel Knowledge Map-Enabled {6D} Movable Antenna Systems With Kinematic Constraints: A Manifold Optimization Approach}, 
	year={2026},
	volume={25},
	number={},
	pages={8968-8981},
	note={early access},
	month={},}

@article{6dma_dis,
	title={{6D} Movable Antenna Enhanced Wireless Network Via Discrete Position and Rotation Optimization},
	author={Shao, Xiaodan and Zhang, Rui and Jiang, Qijun and Schober, Robert},
	journal={IEEE J. Sel. Areas Commun.},
	year={2025},
	volume={43},
	number={3},
	pages={674-687},
	month={Mar.},}
\end{document}